\documentclass[%
 reprint,
 amsmath,amssymb,
 aps,
]{revtex4-2}

\usepackage{graphicx}% Include figure files
\usepackage{dcolumn}% Align table columns on decimal point
\usepackage{bm}% bold math
\usepackage{hyperref}% add hypertext capabilities
\usepackage[all]{hypcap}
\hypersetup{
    colorlinks=true,
    linkcolor=blue,
    }
\usepackage{placeins}
\usepackage{braket}
\usepackage{dsfont, xcolor}
\usepackage{booktabs}
\usepackage{xfrac}
\usepackage[normalem]{ulem}
\usepackage{mathtools}

\newcommand{\zl}[1]{\textcolor{blue}{#1}}

\newcommand{\ba}{\textbf{a}}
\newcommand{\bad}{\textbf{a}^\dagger}
\newcommand{\badd}{\textbf{a}^{\dagger 2}}

\newcommand{\bal}{\bm \alpha}
\newcommand{\bald}{\bm{\alpha}^\dagger}
\newcommand{\baldd}{\bm{\alpha}^{\dagger 2}}
\newcommand{\sgz}{\bm{\sigma}_\textbf{z}}
\newcommand{\sgm}{\bm{\sigma_-}}
\newcommand{\sgp}{\bm{\sigma_+}}
\newcommand{\bd}{\textbf{d}}

\newcommand{\ual}{\underline{\alpha}}

\begin{document}

\preprint{APS/123-QED}

\title{Dynamically enhancing qubit-photon interactions with anti-squeezing}
\author{M. Villiers}
 \email{marius.villiers@ens.fr}
  \affiliation{Laboratoire de Physique de l’Ecole Normale Sup\'erieure, Mines Paris, Centre Automatique et Systèmes,
  Inria, ENS-PSL, Universit\'e PSL, CNRS, Sorbonne Universit\'e, Paris, France}
\author{W.C. Smith}
\email{Present address: Google Quantum AI, Santa Barbara, CA}
 \affiliation{Laboratoire de Physique de l’Ecole Normale Sup\'erieure, Mines Paris, Centre Automatique et Systèmes,
  Inria, ENS-PSL, Universit\'e PSL, CNRS, Sorbonne Universit\'e, Paris, France}
\author{A. Petrescu}
 \affiliation{Laboratoire de Physique de l’Ecole Normale Sup\'erieure, Mines Paris, Centre Automatique et Systèmes,
  Inria, ENS-PSL, Universit\'e PSL, CNRS, Sorbonne Universit\'e, Paris, France}
\author{A. Borgognoni}
 \affiliation{Laboratoire de Physique de l’Ecole Normale Sup\'erieure, Mines Paris, Centre Automatique et Systèmes,
  Inria, ENS-PSL, Universit\'e PSL, CNRS, Sorbonne Universit\'e, Paris, France}
\author{M. Delbecq}%
 \affiliation{Laboratoire de Physique de l’Ecole Normale Sup\'erieure, Mines Paris, Centre Automatique et Systèmes,
  Inria, ENS-PSL, Universit\'e PSL, CNRS, Sorbonne Universit\'e, Paris, France}
 \author{A.~Sarlette}
 \affiliation{Laboratoire de Physique de l’Ecole Normale Sup\'erieure, Mines Paris, Centre Automatique et Systèmes,
  Inria, ENS-PSL, Universit\'e PSL, CNRS, Sorbonne Universit\'e, Paris, France}
 \author{M. Mirrahimi}
 \affiliation{Laboratoire de Physique de l’Ecole Normale Sup\'erieure, Mines Paris, Centre Automatique et Systèmes,
  Inria, ENS-PSL, Universit\'e PSL, CNRS, Sorbonne Universit\'e, Paris, France}
 \author{P. Campagne-Ibarcq}
 \affiliation{Laboratoire de Physique de l’Ecole Normale Sup\'erieure, Mines Paris, Centre Automatique et Systèmes,
  Inria, ENS-PSL, Universit\'e PSL, CNRS, Sorbonne Universit\'e, Paris, France}
\author{T. Kontos}%
 \affiliation{Laboratoire de Physique de l’Ecole Normale Sup\'erieure, Mines Paris, Centre Automatique et Systèmes,
  Inria, ENS-PSL, Universit\'e PSL, CNRS, Sorbonne Universit\'e, Paris, France}
\author{Z. Leghtas}%
 \email{zaki.leghtas@ens.fr}
 \affiliation{Laboratoire de Physique de l’Ecole Normale Sup\'erieure, Mines Paris, Centre Automatique et Systèmes,
  Inria, ENS-PSL, Universit\'e PSL, CNRS, Sorbonne Universit\'e, Paris, France}

\date{\today}

\begin{abstract}
The interaction strength of an oscillator to a qubit grows with the oscillator's vacuum field fluctuations. The well-known degenerate parametric oscillator has revived interest in the regime of strongly-detuned squeezing, where its eigenstates are squeezed Fock states. Owing to the amplified field fluctuations in the anti-squeezed quadrature, it was recently proposed that squeezing this oscillator would dynamically boost qubit-photon interactions. In a superconducting circuit experiment, we observe a two-fold increase in the dispersive interaction between a qubit and an oscillator at 5.5 dB of squeezing, demonstrating in-situ dynamical control of qubit-photon interactions. This work initiates the experimental coupling of oscillators of squeezed photons to qubits, and cautiously motivates their dissemination in experimental platforms seeking enhanced interactions.
\end{abstract}

\maketitle
\section{Introduction}
The magnitude of an electromagnetic oscillator's vacuum field fluctuations sets the scale for its coupling strength to a qubit \cite{Haroche2006}. The value of these fluctuations is directly related to the mode's impedance, and is therefore set by design. For example, the larger the mode impedance, the stronger its electric field will fluctuate, thus enhancing the coupling to the charge degree of freedom of a qubit \cite{Viennot2015, Stockklauser2017, Mi2018, Samkharadze2018}. 
Conversely, the lower the mode impedance, the stronger its magnetic field will fluctuate, thus enhancing the coupling to a spin \cite{Schuster2010, Bienfait2016, Eichler2017}. Despite this design flexibility, some qubits remain difficult to couple to \cite{Cottet2017, Clerk2020}. Recently, Refs.~\cite{Qin2018, Leroux2018} have proposed to boost these fluctuations dynamically. This
would enable in-situ enhancement of qubit-photon interactions,
with far reaching applications such as pushing
weakly coupled systems into the strong coupling regime
\cite{Qin2018, Leroux2018, Lu2015, Lemonde2016, Zeytinoglu2017}, exploring the 
exotic 
ultra-strong regime \cite{Niemczyk2010, Markovic2018, Kockum2019}, and
observing dynamically activated quantum phase transitions \cite{Zhu2020, Shen2022, Chen2021}. 

\begin{figure}[]
\includegraphics[width=\linewidth,keepaspectratio]{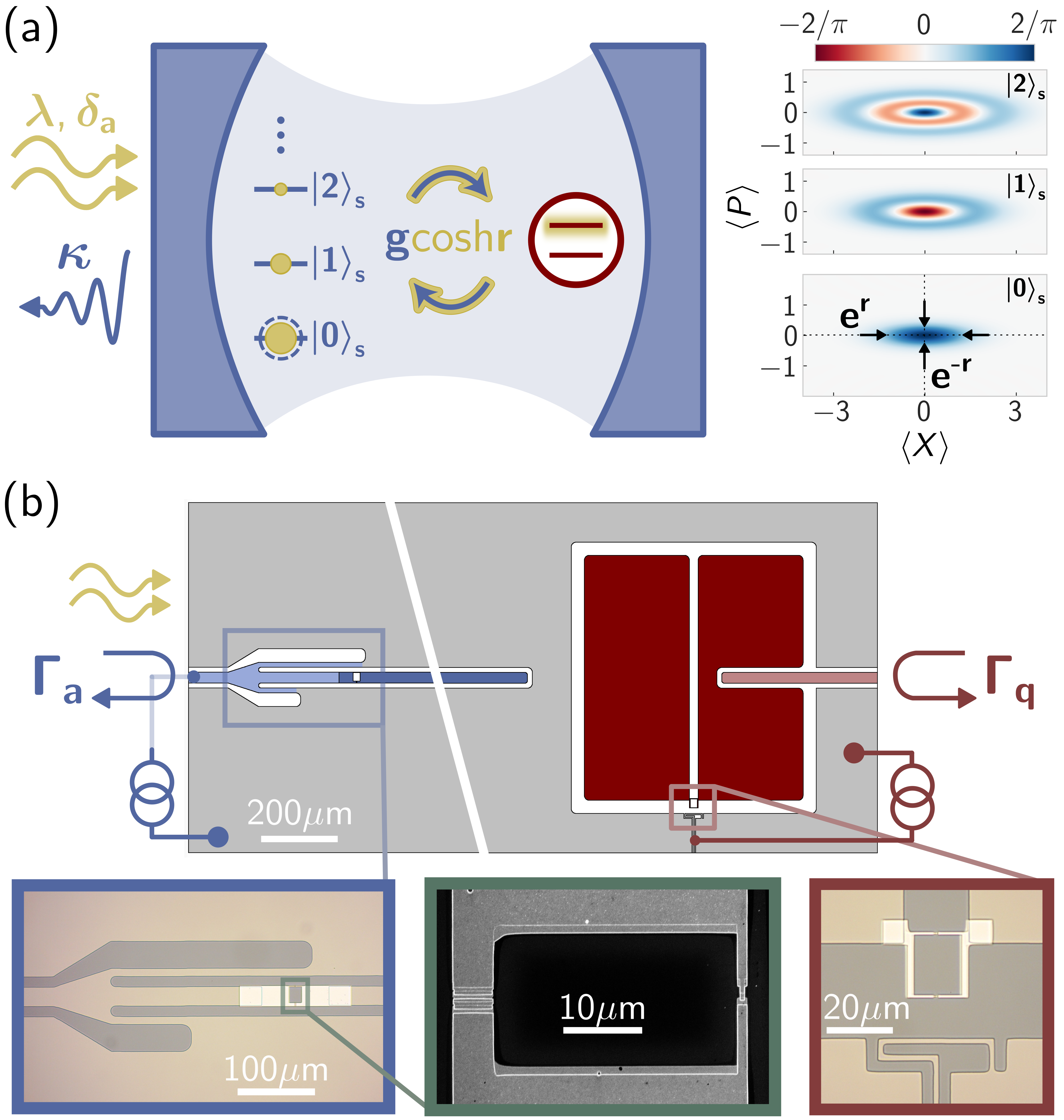}
\caption{Principle of the experiment. (a) A cavity oscillator (blue mirrors) is driven by a two-photon pump with amplitude $\lambda$ and detuning  $\delta_\mathrm{a}$ (gold double arrow) while dissipating at rate $\kappa$ (blue arrow). Its eigenstates (blue levels), that are effectively thermally occupied (full circles), consist of squeezed Fock states $\ket{n}_s$ (right insets: Wigners at $S=e^{2r}=6$~dB). This results in an enhanced coupling $g\cosh{r}$ (gold blue exchange arrows) to a qubit (red levels) and induces qubit dephasing (gold fuzz). (b) Superconducting circuit layout with a diagonal break for compactness. A quarter wavelength coplanar waveguide resonator (blue) implements the oscillator. The reflection spectrum $\Gamma_\mathrm{a}$ is measured through an inductive coupler (optical micrograph in blue inset), that also channels the pump signal and a DC current for flux biasing of the SNAIL loop (scanning electron microscope image in green inset rotated by 90°). The resonator is weakly capacitively coupled to a transmon qubit (bordeaux), which is overcoupled to a transmission line for direct reflection spectroscopy $\Gamma_\mathrm{q}$. A DC line threads flux through its SQUID loop (optical micrograph in bordeaux inset). The oscillator frequency is set to $\omega_\mathrm{a}/2\pi=6.940$~GHz where its decay rate is $\kappa/2\pi=8.7$~MHz. The transmon frequency is set to $\omega_\mathrm{q}/2\pi=6.840$~GHz where its total linewidth is $\gamma_t/2\pi=9.4$~MHz and its charging energy is $E_c/h=114$~MHz. The resonant coupling rate is $g/2\pi=4.9$~MHz (Appendix~\ref{sec:sample}).}
\label{fig:fig1}
\end{figure}

The proposals \cite{Leroux2018, Qin2018} consider a ubiquitous system in quantum optics: the degenerate parametric oscillator (DPO), albeit operated in a new regime (Fig.~\ref{fig:fig1}a). In the usual regime, widely employed for quantum-limited amplification \cite{Castellanos-Beltran2007, Yamamoto2008, Boutin2017, Planat2019, Parker2022, Murch2013, Bienfait2017, Eddins2018}, a pump modulates the oscillator frequency at twice its resonance, thus inducing resonant squeezing. Instead, in the new regime of interest, the pump is far detuned from the parametric resonance \cite{Carmichael1984}. This added detuning renders the system Hamiltonian diagonalizable by a Bogoliubov transformation, and is therefore referred to as a Bogoliubov oscillator (BO) \cite{Metelmann2022}. Unlike a regular harmonic oscillator whose eigenstates are circular Fock states, the eigenstates of a BO are squeezed Fock states. Their amplified fluctuations are the root cause for enhancing qubit-photon interactions. Yet, they also result in enhanced interactions between the BO and its bath, which are expected to dephase the qubit \cite{Shani2022}. Experimentally coupling a BO to qubits was recently achieved with trapped ions, where a phononic BO mediated amplified qubit-qubit interactions, thereby accelerating a M{\o}lmer-S{\o}rensen gate \cite{Burd2021}. Since many physical systems interact through their coupling to an electromagnetic field \cite{Cottet2017, Clerk2020}, photonic BOs have raised high expectations \cite{Leroux2018, Qin2018, Zhu2020, Shen2022, Metelmann2022, Lu2015, Xie2020, Lu2022, Zhong2022}, but have remained experimentally unexplored.

% claim: what? how ?
In a superconducting circuit experiment, we observe that squeezing a BO amplifies its coupling to a qubit. We measure a two-fold increase in the dispersive interaction strength at 5.5 dB of squeezing. Moreover, we demonstrate that BOs -- through enhanced coupling to their bath -- allow for amplification that evades the gain-bandwidth constraint \cite{Metelmann2022}. We observe these phenomena in a Josephson circuit (Fig.~\ref{fig:fig1}b), for which a rich toolbox of nonlinear dipoles is available to couple low loss modes \cite{Blais2021}. While a regular transmon plays the role of the qubit \cite{Koch2007}, implementing a strongly detuned squeezing Hamiltonian without activating spurious nonlinear processes is a technical challenge \cite{Boutin2017, Planat2019}. To this end, we implement the BO by strongly pumping a resonator interrupted by a SNAIL element \footnote{SNAIL: superconducting nonlinear asymmetric inductive element}, capable of inducing significant squeezing while maintaining a vanishingly small Kerr 
non-linearity 
\cite{Frattini2018, Sivak2019}.

\section{Theory}
A SNAIL-resonator with bare frequency $\omega_\mathrm{a}$, pumped at a frequency $\omega_\mathrm{p}$ detuned from the degenerate parametric resonance $2\omega_\mathrm{a}$, emulates the DPO model in a frame rotating at half the pump frequency (Appendix~\ref{sec:theory_dpo}). 
It is described by the following Hamiltonian and Lindblad operator:
\begin{equation}
\label{eq:Ha}
\begin{aligned}
    \bm{\mathcal{H}}_\textbf{ph}/\hbar &= \delta_\mathrm{a} \ba^\dagger\ba - \frac{\lambda}{2}\left(\ba^2 + \ba^{\dagger 2} \right)\:,\;\;
    \bm{L}_\textbf{ph}=\sqrt{\kappa}\ba\:,
\end{aligned}
\end{equation}
where $\ba$ is a bosonic annihilation operator, $\delta_\mathrm{a}$ is the detuning between the oscillator and half the pump frequency, $\lambda$ is the amplitude of the two-photon pump, and $\kappa$ is the dissipation rate.
This system is widely operated in the resonant squeezing regime $\delta_\mathrm{a}=0$ and $\lambda<\kappa/2$, for near-quantum limited amplification and squeezed radiation generation \cite{Castellanos-Beltran2007, Yamamoto2008, Boutin2017, Planat2019, Parker2022, Murch2013, Bienfait2017, Eddins2018}. Interestingly, at $\delta_\mathrm{a}=0$, the dynamics would be unstable in absence of dissipation.

Instead, we focus on the detuned squeezing regime $\kappa/2\ll\lambda<|\delta_\mathrm{a}|$. Introducing the squeezing parameter $r$ such that $\tanh2r=\lambda/|\delta_\mathrm{a}|$ and the squeezing amplitude $S=e^{2r}$
, we diagonalize Hamiltonian~\eqref{eq:Ha} through the Bogoliubov transformation $\bm{\mathcal{U}_s}=e^{\frac{r}{2}(\ba^2-{\ba^\dagger}^2)}$. We introduce the canonical Bogoliubov operator $\bal\equiv \bm{\mathcal{U}_s}^\dag\ba \bm{\mathcal{U}_s}$ so that the Hamiltonian and Lindblad operator~\eqref{eq:Ha} rewrite:
\begin{equation}
\label{eq:Halpha}
\begin{gathered}
    \bm{\mathcal{H}}_\textbf{ph}/\hbar = \Omega_\mathrm{a}[r] \bal^\dagger\bal\;,\; \\
    \bm{L}_\textbf{ph}=\sqrt{\kappa}\left(\bal \cosh r+\bal^\dagger\sinh r\right)\;,
\end{gathered}
\end{equation} 
where $\Omega_\mathrm{a}[r]=\delta_\mathrm{a}/\cosh2r$, and the Lindblad operator depicts a squeezed bath which, in the limit $|2\Omega_\mathrm{a}[r]|\gg\kappa\cosh2r$, effectively amounts to a thermal bath with occupancy $\sinh^2r$ \cite{Lemonde2016,Villiers2023}. The eigenstates of $\bm{\mathcal{H}}_\textbf{ph}$ are squeezed Fock states $\ket{n}_s\equiv\bm{\mathcal{U}_s}^\dag\ket{n}$, where $\ket{n}$ is the $n^{th}$ Fock state of mode $\ba$. For every eigenstate, the fluctuations of the squeezed quadrature are reduced by a factor $e^{-r}$, and inversely, those of the anti-squeezed quadrature are enhanced by $e^{r}$ (Fig.~\ref{fig:fig1}a). We stress that the squeezing of these eigenstates, that is different from the squeezing of a steady-state (Appendix~\ref{sec:ss_squeezing}), is the root cause of the enhanced coupling of a BO to other modes. This applies to the continuum of bath modes, with striking applications for broadband quantum limited amplification \cite{Metelmann2022}. It also applies to a discrete mode of interest such as a qubit.

\begin{figure*}[]
    \includegraphics[width=\linewidth,keepaspectratio]{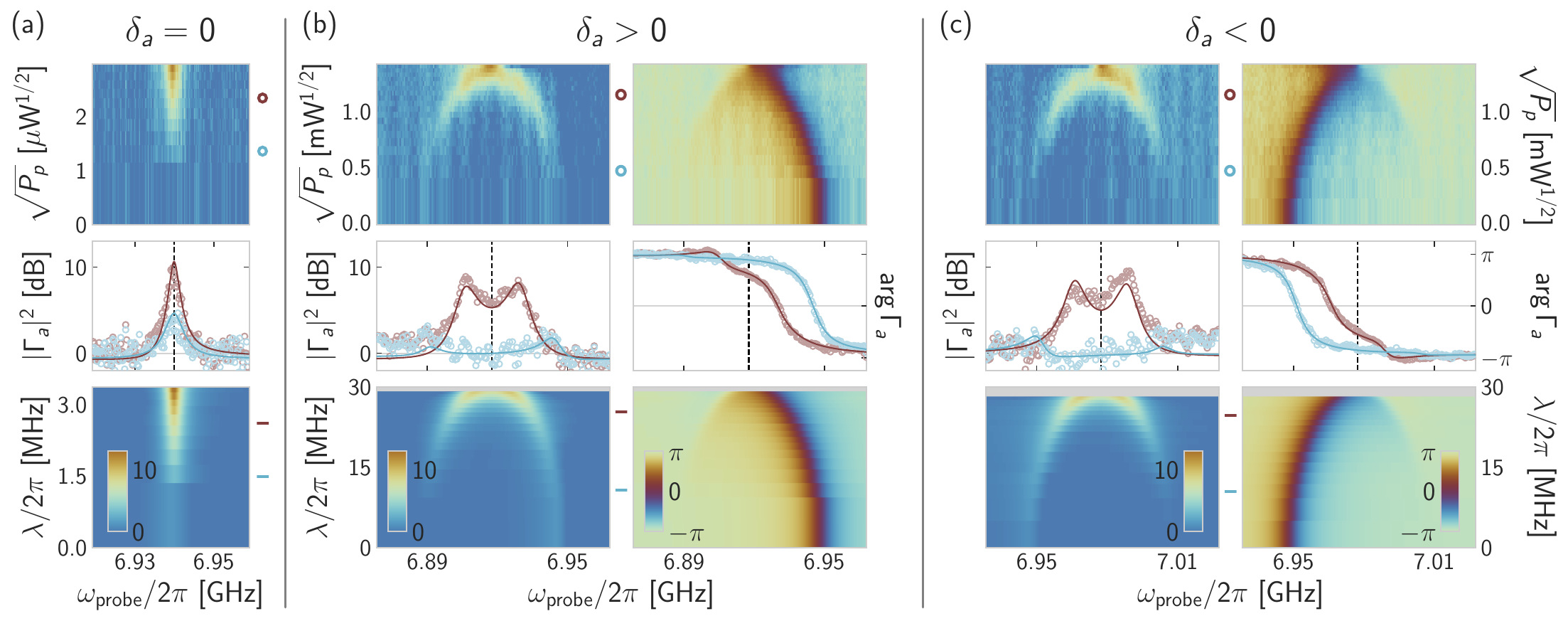}
    \caption{\label{fig:wide} Microwave response of the Bogoliubov oscillator. (a) The pump frequency is set to $\omega_\mathrm{p}=2\omega_\mathrm{a}$. Top: Reflection gain (color) versus probe frequency (x-axis) and the square root of the pump power at 300~K (y-axis). The colorbar (in decibels) is indicated in the bottom panel. Bottom: fitted reflection gain from input-output theory applied to Eq.~\eqref{eq:Ha}, with $\lambda$ as the only free parameter (Appendix~\ref{sec:dpo_cal}). Middle: line-cuts of the measured (open circles) and fitted (solid lines) gain (y-axis) versus probe frequency (x-axis). Colored dots and segments in the top and bottom panels indicate the location of the line-cuts. The vertical dashed line marks half the pump frequency. (b,c) The pump frequency is set to  $\omega_\mathrm{p}=2\omega_\mathrm{a}- 2\delta_\mathrm{a}$, where $\delta_\mathrm{a}/2\pi=\pm30$~MHz. The reflection gain response (left) follows the same procedure as (a). Also represented is the reflection phase response (right). The data for $\delta_\mathrm{a}/2\pi=\pm 20, \pm40$~MHz (not represented) display the same features.}
    \label{fig:SPA_squeeze}
\end{figure*}

In this experiment we couple our BO to a transmon qubit \cite{Koch2007}. For clarity the following theory is derived for a two-level system, while the higher transmon levels are accounted for when fitting the data (Appendix~\ref{sec:theory_trm}). The Hamiltonian of the coupled system in the rotating frame is $\bm{\mathcal{H}}_\textbf{q-ph} = \hbar\delta_\mathrm{q}\frac{\sgz}{2}  + \bm{\mathcal{H}}_\textbf{ph} + \hbar g\left( \ba \sgp + \ba^\dagger\sgm \right),$
where $\delta_\mathrm{q}$ is the detuning of the qubit to half the pump frequency, $g$ the interaction strength and $(\sgp, \sgm, \sgz)$ are the raising, lowering and Pauli $Z$ operators. The Lindblad operators associated to the qubit relaxation and dephasing are $\bm{L_}-=\sqrt{\gamma_1}\sgm$, $\bm{L_\phi}=\sqrt{\gamma_\phi/2}\sgz$.
In the Bogoliubov basis, the Hamiltonian reads:
\begin{align}
\label{eq:HRabi}
    \bm{\mathcal{H}}_\textbf{q-ph}/\hbar =& \:\Omega_\mathrm{a}[r] \bald\bal +\delta_\mathrm{q}\frac{\sgz}{2} \\
    +\:g\cosh r&\left(\bal\sgp + \bald\sgm \right) +g\sinh r\left(\bal\sgm + \bald\sgp \right)\;. \nonumber
\end{align}
The enhanced interaction strength is immediately visible in Eq.~\eqref{eq:HRabi} where $g$ is multiplied by $\cosh{r}$ [resp: $\sinh{r}$] for the excitation number conserving [resp: non-conserving] terms. Such an enhancement is neither expected when injecting squeezed radiation on the oscillator \cite{Murch2013, Bienfait2017, Eddins2018}, as it does not change the system Hamiltonian, nor when the squeezing is resonant \cite{Eddins2019}, as it does not lead to squeezed Fock states as eigenstates. On resonance $\delta_\mathrm{q}=\Omega_\mathrm{a}[r]$, and provided $|2\Omega_\mathrm{a}[r]| \gg g\sinh r$, the system reduces to an enhanced resonant Jaynes-Cummings interaction, that could be unambiguously revealed through an increased vacuum-Rabi splitting \cite{Villiers2023}. However, this simple picture is blurred by the squeezed bath that populates higher BO energy levels \cite{Ong2013, Bishop2009, Bonsen2022}, which in turn broadens the qubit spectral line \cite{Shani2022} (Fig.~\ref{fig:fig1}a). Refs.~\cite{Leroux2018, Qin2018} propose to circumvent this problem by injecting orthogonally squeezed vacuum, so that the bath viewed by the BO remains in vacuum. While possible in principle, injecting squeezed vacuum is easily contaminated by damping in transmission lines and cavity internal losses, thereby remaining a technical challenge \cite{Murch2013, Bienfait2017, Eddins2018}.

Instead, the dispersive regime is well adapted to measuring the qubit-BO coupling through qubit spectroscopy \cite{Blais2004, Schuster2005, Gambetta2006}, even in the presence of a non-vanishing bath occupation \cite{Schuster2007, Ong2011, Viennot2018, Dassonneville2021}. We place ourselves in the dispersive regime where $ge^r, \kappa, \gamma_1, \gamma_\phi$, when divided by $|\delta_\mathrm{q}\pm\Omega_\mathrm{a}[r]|$, are of order $\eta\ll 1$. Note that we require the qubit to be detuned from both the BO resonance and its mirror idler frequency. Moreover, we chose an intermediate coupling regime $g\sim \gamma_1 \sim \kappa$ in order to mimic systems that could readily benefit from enhanced couplings \cite{Cottet2017, Clerk2020, Vine2023}. Retaining up to second order terms in $\eta$, the loss operators remain unchanged and Hamiltonian~\eqref{eq:HRabi} is diagonalized as (Appendix~\ref{sec:theory_qubit}):
\begin{equation}
\label{eq:Hchi}
    \bm{\mathcal{H}}_\textbf{q-ph}/\hbar = \Omega_\mathrm{a}[r]\bald \bal 
    + \left[ \delta_\mathrm{q}  + \chi[r] \Big(\bald \bal + \frac{1}{2} \Big)\right] \frac{\sgz}{2} \;,
\end{equation}
where:
\begin{equation}
\label{eq:chi}
\chi[r] = \frac{2g^2}{\delta_\mathrm{q}-\Omega_\mathrm{a}[r]}\cosh^2r + \frac{2g^2}{\delta_\mathrm{q}+\Omega_\mathrm{a}[r]}\sinh^2r \;.
\end{equation}
The enhanced dispersive coupling is immediately visible in Eq.~\eqref{eq:chi}. The first term has the familiar form of a dispersive interaction with a modified detuning and a $g$ coupling enhanced by $\cosh{r}$, while the second term emerges from the presence of the mirror idler resonance. Observing the dispersive coupling enhancement is the main goal of this experiment.

\section{The Bogoliubov oscillator}
Parametric oscillators have long been employed to amplify signals for qubit readout and generate squeezed radiation \cite{Castellanos-Beltran2007, Yamamoto2008, Boutin2017, Planat2019, Bienfait2017, Eddins2018}. In our three-wave mixing device, we observe amplification by setting the pump frequency at the parametric resonance $\omega_\mathrm{p}=2\omega_\mathrm{a}$ (Fig.~\ref{fig:SPA_squeeze}a). By increasing the pump power close to the parametric instability $\lambda= \kappa/2$, we observe up to 12 dB of gain. We enter the regime of the BO by detuning the pump away from the parametric resonance $\omega_\mathrm{p}=2\omega_\mathrm{a}-2\delta_\mathrm{a}$, where the detuning verifies $|\delta_\mathrm{a}|\gg\kappa/2$. When $\delta_\mathrm{a}/2\pi=+30$~MHz (Fig.~\ref{fig:SPA_squeeze}b), as we increase the pump power, the oscillator resonance shifts down from $\omega_\mathrm{a}$ to $\omega_\mathrm{p}/2$, following the theoretical prediction $\omega_\mathrm{p}/2 + \Omega_\mathrm{a}[r]$ where $\Omega_\mathrm{a}[r]=\delta_\mathrm{a}/\cosh{2r}$. Moreover, this resonator of squeezed photons responds to regular plane waves at a mirror frequency $\omega_\mathrm{p}/2-\Omega_\mathrm{a}[r]$. This idler peak merges into the signal peak when $\lambda\gtrsim\sqrt{\delta_\mathrm{a}^2-\kappa^2/4}$, which we refer to as the coalescent regime \cite{Villiers2023}. Symmetrically, for $\delta_\mathrm{a}/2\pi=-30$~MHz (Fig.~\ref{fig:SPA_squeeze}c), the oscillator resonance shifts up from $\omega_\mathrm{a}$ to $\omega_\mathrm{p}/2$. This symmetric behavior differs from the response of a Kerr oscillator to a detuned pump, where the sign of the Kerr sets the  direction of the shift, independently of the pump frequency. The results of Fig.~\ref{fig:wide} demonstrate that $\lambda$ -- the only fit parameter relating data and theory -- is reliably identified at every pump power, thereby fully characterizing the BO and the squeezing of its eigenstates.

\begin{figure}[]
\includegraphics[width=\linewidth,keepaspectratio]{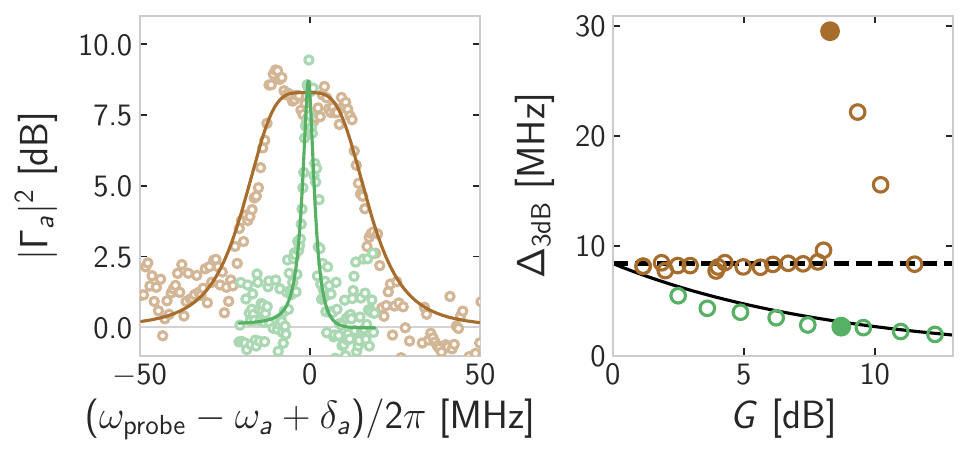}
\caption{The Bogoliubov amplifier: comparison of the amplification bandwidth at $\delta_\mathrm{a}=0$ and $|\delta_\mathrm{a}|\gg\kappa/2$. Left: reflection gain (y-axis) versus input signal frequency (x-axis). For a similar maximum gain of about 9 dB, the amplification bandwidth for $\delta_\mathrm{a}/2\pi=+30$~MHz (brown) is much larger than for $\delta_\mathrm{a}=0$ (green). The data (open circles) are fitted (solid lines) from input-output theory applied to Eq.~\eqref{eq:Ha} (Appendix~\ref{sec:dpo_cal}). Right: 3 dB amplification bandwidth (y-axis) versus maximum gain $G=\mathrm{max}_\omega|\Gamma_\mathrm{a}(\omega)|^2$ (x-axis).  The data (open circles) for $\delta_\mathrm{a}=0$ (green) follow the typical constant gain-bandwidth product trend (solid line). Setting $\delta_\mathrm{a}/2\pi=+30$~MHz (brown), a constant bandwidth is maintained  (dashed line) until the signal and idler peaks merge. }
\label{fig:SPA_gbw}
\end{figure}

A striking feature appears in the amplitude response of the oscillator, where gain is observed at both signal and idler frequencies. Indeed, in the resonant regime $\delta_\mathrm{a}=0$, the 3~dB amplification bandwidth $\Delta_\text{3dB}$ reduces with gain $G$ according to the gain-bandwidth product constraint $\Delta_{3\text{dB}}\sqrt{G}=\kappa$ (Fig.~\ref{fig:SPA_gbw}). In contrast in the detuned regime, following either the signal or idler peak, we observe a constant amplification bandwidth, independently of the gain. As demonstrated in Appendix~\ref{sec:theory_inout}, this fixed bandwidth can be traced back to the enhanced interactions of the BO to its bath modes. Finally in the coalescent regime, the two peaks merge and the amplification bandwidth more than doubles. This amplifier, praised for evading the fundamental gain-bandwidth constraint, has been coined the Bogoliubov amplifier \cite{Metelmann2022}.

\FloatBarrier

\section{Qubit spectroscopy in the presence of squeezed photons} 

\begin{figure}
\includegraphics[width=\linewidth,keepaspectratio]{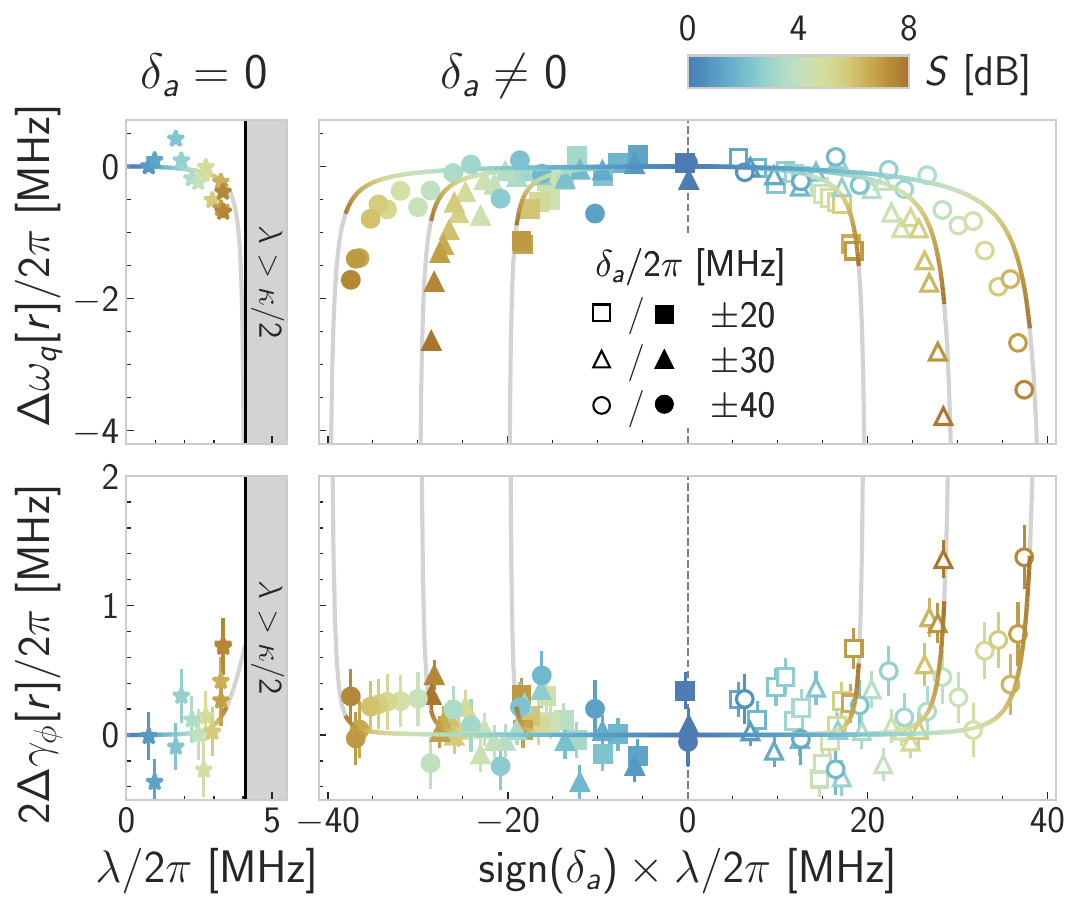}
\caption{Spectral response of a qubit interacting with squeezed photons. Left: The pump frequency is set to $\omega_\mathrm{p}=2\omega_\mathrm{a}$. Top (Bottom): qubit frequency shift (induced dephasing) versus pump amplitude (x-axis). The data (stars) are fitted with analytical expressions (solid lines) adapted from \cite{Eddins2019} (Appendix~\ref{sec:disp_amp}). The shaded area marks the instability region. Right: Same as left panels with the pump frequency set to $\omega_\mathrm{p} = 2\omega_\mathrm{a} - 2\delta_\mathrm{a}$, where $\delta_\mathrm{a}/2\pi \in \{\pm20,\pm30,\pm40\}$ MHz. The solid lines correspond to Eqs.~\eqref{eq:spectral} adapted for a transmon (Appendix~\ref{sec:theory_trm}). A common colormap maps the pump amplitude to an equivalent squeezing (steady-state anti-squeezing, see Appendix~\ref{sec:ss_squeezing}) for $\delta_\mathrm{a}\neq0$ ($\delta_\mathrm{a}=0$) in decibels.}
\label{fig:TRMsqueeze}
\end{figure}

After having characterized the BO, we now turn to the impact of its amplified fluctuations on the qubit. The oscillator, whose eigenstates are now squeezed Fock states, is expected to couple more strongly not only to its bath but also to the qubit. Both aspects will affect the qubit spectrum. In the weak dispersive regime $|\chi[r]|\ll\kappa$, we relate the qubit frequency shift $\Delta\omega_\mathrm{q}[r]$ and induced dephasing $\Delta\gamma_\phi[r]$ to the mean and correlation function of the BO number operator. We find (Appendix~\ref{sec:theory_mid}):
\begin{equation}
\begin{aligned}
    \Delta \omega_\mathrm{q}[r] &= \chi[r]\bigg(\frac{1}{2}+\sinh^2r\bigg)-\frac{1}{2}\chi[0]\;, \\
    \Delta\gamma_\phi[r] &= \frac{\chi^2[r]}{\kappa} \left(1+\sinh^2r\right)\sinh^2r\;,
    \label{eq:spectral}
\end{aligned}
\end{equation}
where $\chi[r]$ is the modified interaction strength given by Eq.~\eqref{eq:chi}, and $\sinh^2r$ the mean occupancy of the BO mode resulting from its modified bath coupling. These equations are derived for a two level system and are adapted for a transmon to fit our data (Appendix~\ref{sec:theory_trm}).
The frequency shift $\Delta \omega_\mathrm{q}[r]$ can be decomposed in two parts. First, a photon number independent term $\frac{1}{2}\chi[r]$, which is reminiscent of the Lamb shift experienced by an atom immersed in the vacuum fluctuations of an electromagnetic mode. Second, a photon number dependent term $\chi[r]\sinh^2r$, which is reminiscent of the AC-Stark effect. The term $\chi[0]/2$ is subtracted since the frequency shift is referenced to the absence of pump ($r=0$). Interestingly, the expression of the induced dephasing $\Delta\gamma_\phi[r]$ is akin to the dephasing of a qubit dispersively coupled to a mode of thermal occupation $\sinh^2r$ \cite{Bertet2005, Rigetti2012}. This reveals that the qubit experiences the squeezed bath populating higher BO energy levels, as a thermal bath. In principle, this induced decoherence, flagged by \cite{Shani2022}, could be cancelled by injecting conversely squeezed radiation while preserving the interaction enhancement \cite{Qin2018, Leroux2018}.

The qubit-oscillator detuning is set to $-100$~MHz, thus placing the system in the weak dispersive regime $\chi[r=0]/2\pi=-250$~kHz (Appendix~\ref{sec:disp_cal}). For various pump detunings $\delta_\mathrm{a}/2\pi\in\{0,\pm20,\pm30,\pm40\}$~MHz, and pump amplitudes inducing up to 8 dB of squeezing $S$, we acquire the qubit reflection spectrum through its dedicated port. From each spectrum, we extract the frequency shift $\Delta \omega_\mathrm{q}[r]$ and linewidth broadening $2\Delta\gamma_\phi[r]$ referenced to $S=0$~dB (pump off) (Fig.~\ref{fig:TRMsqueeze}). This technique was preferred to a time-resolved analysis as a consequence of the large transmon linewidth (Appendix~\ref{sec:transmon_spec}). For $\delta_\mathrm{a}=0$, the balance of resonant two-photon pumping and dissipation stabilizes a squeezed steady-state. At maximal steady-state anti-squeezing, the oscillator mean occupancy is found to be of less than 2 photons (Appendix~\ref{sec:ss_squeezing}). Hence, the variations of $\Delta \omega_\mathrm{q}[r]$ and $\Delta\gamma_\phi[r]$ are consistent with a constant dispersive interaction strength (see Fig.~\ref{fig:TRMsqueeze} left). This is in stark contrast with the case $|\delta_\mathrm{a}|\gg \kappa/2$, where the two-photon pump is balanced, not by dissipation, but by the detuning $\delta_\mathrm{a}$. Three notable features are visible in Fig.~\ref{fig:TRMsqueeze} right. First, for each detuning, as the pump amplitude approaches the instability point $\lambda=|\delta_\mathrm{a}|$ where the squeezing parameter $r=\frac{1}{2}\tanh^{-1}{\lambda/|\delta_\mathrm{a}|}$ diverges, we observe rapidly increasing frequency shifts and line broadenings. Second, the symmetry between positive and negative detunings is broken. Indeed, the BO frequency shifts towards the qubit for $\delta_\mathrm{a}>0$ and away from the qubit for $\delta_\mathrm{a}<0$. Interestingly, despite this asymmetry, the qubit frequency shifts down with increasing $\lambda$, regardless of the sign of $\delta_\mathrm{a}$, showing that the dominant effect at play is the BO enhanced fluctuations, and not a trivial modulation of the BO-qubit detuning. Finally, the magnitudes of the qubit spectral shift and broadening are large. At maximal squeezing, the qubit frequency shifts by at least 4 times the bare qubit-BO dispersive coupling. Such large shifts cannot be explained by an unchanged interaction strength and a simple increase in the BO population. Indeed, we estimate $\sinh^2r\le 1.2$ over this entire data-set, thus hinting towards a significant enhancement of the qubit-BO interaction strength.

\section{Enhancing the dispersive interaction via anti-squeezing}

\begin{figure}
    \includegraphics[width=\linewidth,keepaspectratio]{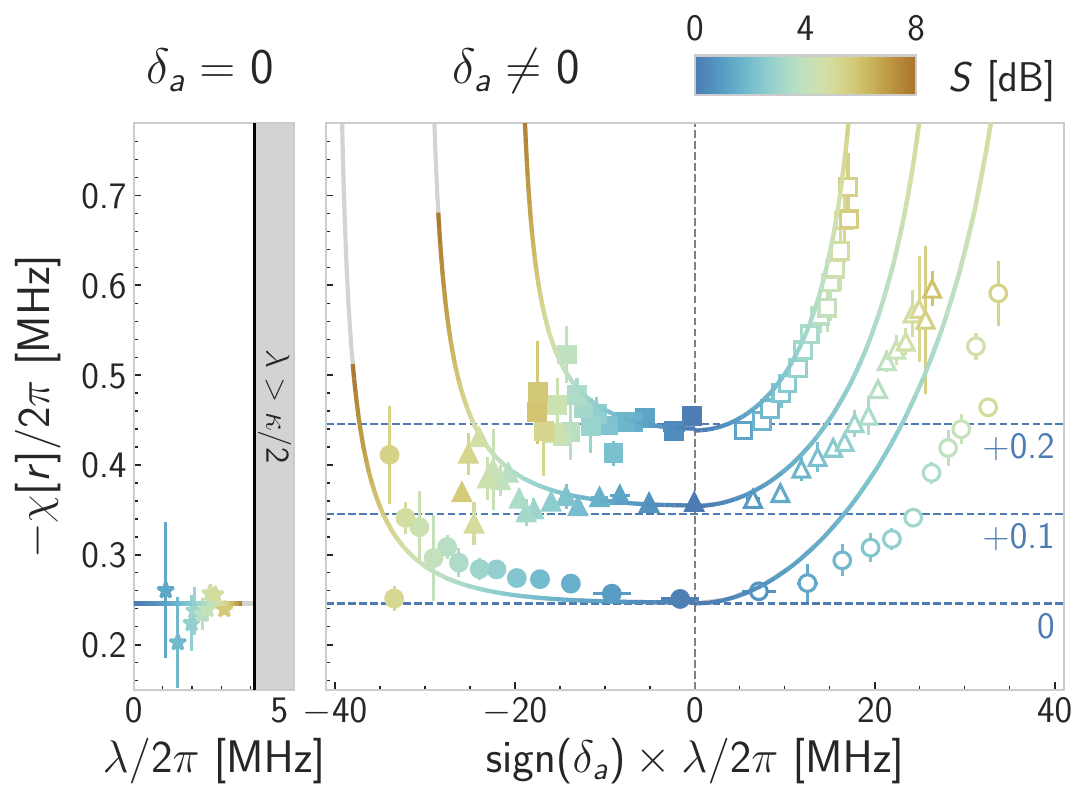}
    \caption{Enhanced dispersive interaction of a qubit with squeezed photons. Left: the pump frequency is set to $\omega_\mathrm{p}=2\omega_\mathrm{a}$. The dispersive interaction strength (y-axis) does not increase with the two-photon pump amplitude (x-axis). The solid line marks the bare dispersive coupling. Right: same as left panel for $\delta_\mathrm{a}/2\pi\in\{\pm20,\pm30,\pm40\}$~MHz, where families of same $|\delta_\mathrm{a}|$ (squares, triangles, circles) are offset by $\{0.2, 0.1, 0\}$~MHz for clarity (empty symbols for $\delta_\mathrm{a}>0$, full symbols for $\delta_\mathrm{a}<0$). Solid lines correspond to the transmon version of Eq.~\eqref{eq:chi} (Appendix~\ref{sec:theory_trm}). In the BO regime, the dispersive interaction strength increases notably with squeezing. A common colormap maps the pump amplitude to an equivalent squeezing (steady-state anti-squeezing, \zl{see Appendix~\ref{sec:ss_squeezing}}) for $\delta_\mathrm{a}\neq0$ ($\delta_\mathrm{a}=0$) in decibels.}
    \label{fig:chi_vs_S}
\end{figure}

We measure the dispersive coupling $\chi[r]$ by adapting the procedure of Refs.~\cite{Schuster2005, Gambetta2006} (Fig.~\ref{fig:chi_vs_S}). For each pump detuning $\delta_\mathrm{a}$ and amplitude $\lambda$, we apply a weak drive tone on the BO at its frequency $\frac{\omega_\mathrm{p}}{2}+\Omega_\mathrm{a}[r]$. The drive power is quantified in units of photon number $\bar{n}_\mathrm{d}$ that would be injected in the oscillator in the absence of squeezing (Appendix~\ref{sec:disp_cal}). For each value of $\bar{n}_\mathrm{d}$, a qubit reflection spectrum is acquired. Two effects are observed: a frequency shift $\Delta\omega_\mathrm{q}[r,\bar{n}_\mathrm{d}]$ and a linewidth broadening $\Delta\gamma_\phi[r,\bar{n}_\mathrm{d}]$. Adapting the procedure that resulted in Eq.~\eqref{eq:spectral} in the presence of the drive far from coalescence ($|\Omega_\mathrm{a}[r]|\gg \kappa/2$), we find (Appendix~\ref{sec:theory_mid}):
\begin{equation}
    \begin{aligned}
        \Delta\omega_\mathrm{q}[r,\bar{n}_\mathrm{d}] &= \chi[r] \bar{n}_\mathrm{d}\cosh^2r\;, \\
        \Delta\gamma_\phi[r,\bar{n}_\mathrm{d}] &= \frac{2\chi^2[r]}{\kappa} \big(1+2\sinh^2r\big) \bar{n}_\mathrm{d}\cosh^2r\;.
    \end{aligned}
\end{equation}
Both quantities are referenced to $\bar{n}_\mathrm{d}=0$. 
The frequency shift resembles the usual AC-Stark shift albeit the extra term $\cosh^2r$ that indicates the enhanced coupling of the BO to the incident drive. On the other hand, the linewidth broadening has a form corresponding to the induced dephasing by an amplified coherent drive $\bar{n}_d\cosh^2r$, superimposed to the effective thermal occupation $\sinh^2r$. We fit the measured frequency shifts and linewidth broadenings to these derived expressions, keeping only $\chi[r]$ as a free parameter, and report the results in Fig.~\ref{fig:chi_vs_S}.

The left panel of Fig.~\ref{fig:chi_vs_S} displays a control experiment at $\delta_\mathrm{a}=0$ that shows no enhancement in $\chi$ as expected by theory (Appendix~\ref{sec:disp_amp}). The right panel displays $\chi[r]$ versus $\lambda$ in the BO regime $|\delta_\mathrm{a}|\gg \kappa/2$. As previously observed in Fig.~\ref{fig:TRMsqueeze}, the symmetry between positive and negative detunings is broken. This is expected since two different effects contribute to the variation of $\chi[r]$ with squeezing. First, the enhanced fluctuations of the BO result in an enhanced interaction strength, revealed by the $\cosh^2r$, $\sinh^2r$ factors in Eq.~\eqref{eq:chi}. This effect is independent of the sign of $\delta_\mathrm{a}$. Second, as the BO is squeezed, its frequency $\Omega_\mathrm{a}[r]$ varies thus modifying the qubit-BO detuning. It is this effect that depends on the sign of $\delta_\mathrm{a}$. For positive pump detunings (empty symbols), the BO shifts towards the qubit so the two contributions add, resulting in a significant increase in $\chi[r]$. We measure up to a two-fold increase in $\chi[r]$ for $\delta_\mathrm{a}/2\pi=+20$~MHz, from $\chi[r=0]/2\pi=-250$~kHz to $\chi[r]/2\pi=-510$~kHz at $\lambda/2\pi=17$~MHz corresponding to $S=5.5$~dB of squeezing. Only 15\% of this increase is attributed to the reduced qubit-BO detuning. The converse is true for negative pump detunings (full symbols): the BO moves away from the qubit. Remarkably, the effect of enhanced fluctuations outweighs the effect of increased detuning, resulting in a measurable, yet modest, increase in $\chi[r]$ even for negative detunings. The matching of theory to data noticeably degrades at large $+|\delta_\mathrm{a}|$, possibly due to the narrowing proximity of the idler peak to the qubit.

\section{Conclusion}
In conclusion, we have observed a two-fold increase in the dispersive interaction between a qubit and a BO at 5.5~dB of squeezing. A word of caution is however necessary. The BO, through its amplified field fluctuations, couples more strongly not only to the qubit but to all coupled modes, including its bath. The resulting finite occupation of the BO induces decoherence on the qubit, as warned by Ref.~\cite{Shani2022} and observed in this experiment. Future experiments could evade this induced decoherence by conversely squeezing the bath modes, as proposed by Refs.~\cite{Leroux2018,Qin2018}. In practice, it would involve injecting squeezed vacuum through the transmission line coupled to the BO \cite{Murch2013,Bienfait2017,Eddins2018}. The squeezing amplitude and phase would be tuned to maintain the BO close to its vacuum state, while preserving the enhanced qubit-photon interactions.

Yet, the future of BOs is not prescribed to the injection of squeezed vacuum. Regarding the BO alone, the new regime of amplification that evades the gain-bandwidth product constraint \cite{Metelmann2022} is immediately applicable for broadband quantum limited amplification with no hardware overhead \cite{Vine2023}. As for applications to enhance couplings, the BO is readily applicable to systems with weak couplings to photons and strong intrinsic dephasing so that the squeezing-induced dephasing is marginal. Common examples are quantum dots \cite{Cottet2017} and spin ensembles \cite{Clerk2020}. This enhanced coupling could then be leveraged by the BO gain for improved qubit readout \cite{Eddins2019}. Finally, the ability to dynamically tune qubit-photon interactions comes as a great resource for the study of squeezing-induced phase transitions \cite{Zhu2020,Shen2022,Chen2021}, quantum transduction \cite{Zhong2022}, and the exploration of the ultra-strong coupling regime \cite{Niemczyk2010,Markovic2018,Kockum2019}.

\FloatBarrier
\paragraph*{Author contributions}
{M.V, T.K and Z.L conceived the experiment. M.V designed the sample with guidance from W.C.S. M.V fabricated and measured the sample. M.V and Z.L analyzed the data and wrote the manuscript with input from all authors. M.V, A.P and Z.L derived the theory with support from P.C-I, A.S and M.M. Experimental support was provided by W.C.S, A.B, M.D and T.K.}

\begin{acknowledgments}
We thank R. Assouly, R. Dassonneville, E. Flurin and R. Lescanne for fruitful discussions. We thank Lincoln Labs for providing a Josephson Traveling-Wave Parametric Amplifier. The devices were fabricated within the consortium Salle Blanche Paris Centre. We thank Jean-Loup Smirr and the Collège de France for providing nano-fabrication facilities. This work was supported by the QuantERA grant QuCOS, by ANR 19-QUAN-0006-04. Z.L.\ acknowledges support from ANR project ENDURANCE, and EMERGENCES grant ENDURANCE of Ville de Paris. This work has been supported by the Paris \^{I}le-de-France Region in the framework of DIM SIRTEQ. This project has received funding from the European Research Council (ERC) under the European Union’s Horizon 2020 research and innovation programme grant agreements No.\ 851740.
\end{acknowledgments}

%\end{document}

%\clearpage
\appendix

\section{Sample and setup}
\label{sec:sample}
\subsection{Circuit implementation}
We implement a BO coupled to a qubit in a circuit quantum electrodynamics (cQED) coplanar waveguide architecture (Fig.~\ref{fig:fig1}b). The oscillator is fabricated from a quarter wavelength superconducting resonator shunted to ground through a superconducting nonlinear asymmetric inductive element (SNAIL) element \cite{Frattini2018}. This element consists of three large Josephson junctions in parallel with a small one, forming a loop threaded by magnetic flux. The SNAIL endows the resonator with non-linearity that has a vanishing Kerr at a well chosen flux, while maintaining a significant three-wave mixing term \cite{Sivak2019}. This choice of non-linear element was essential to implement Hamiltonian \eqref{eq:Ha} with minimal parasitic terms (Appendix~\ref{sec:theory_dpo}). An inductive coupler channels both direct current (DC) for flux biasing, and radio-frequency (RF) probe and pump tones \cite{Bothner2013, Besedin2018}. At the Kerr-free point, the oscillator frequency is $\omega_\mathrm{a}/2\pi=6.940$~GHz, and its dissipation rate $\kappa/2\pi=8.7$~MHz, largely dominated by coupling to the transmission line (Appendix~\ref{sec:dpo_cal_kerr_free}). The resonator is capacitively coupled to a flux tunable transmon. The transmon is coupled to a transmission line for direct reflection spectroscopy, inducing a total linewidth $\gamma_t/2\pi=9.4$~MHz at $\omega_\mathrm{q}/2\pi=6.840$~GHz (Appendix~\ref{sec:transmon_resonant}). 
%The relaxation rate $\gamma_1$ and dephasing rate $\gamma_\phi$ are related to the total linewidth through: $\gamma=\gamma_1+2\gamma_\phi$. 
Since the transmon anharmonicity $E_c/h=114$~MHz is much larger than $\gamma_t$, its two lowest energy eigenstates implement the qubit (Appendix~\ref{sec:transmon_twotone}). The resonant coupling strength is $g/2\pi=4.9$~MHz (Appendix~\ref{sec:transmon_resonant}).
%This experiment is designed to simulate coupling enhancement to weakly coupled systems such as spin qubits \cite{Viennot2015}, \zl{therefore the regime of interest is the weak dispersive regime where the coupling strength $g/2\pi=4.9$~MHz induces a dispersive coupling smaller than the mode dissipation rates.}

\subsection{Fabrication}
\label{fabrication}
The sample is made out of a 280~$\mu$m thick intrinsic silicon chip, sputtered with 100~nm of niobium. A first laser lithography step patterns the large features of the circuit on S1805 resist. It is revealed in MF319, and subsequently etched with SF$_6$. The Al/AlOx/Al Josephson junctions are fabricated during a second step of electronic lithography, using a Dolan bridge technique on a bilayer of MMA/MAA and PMMA. After reveal in a 1:3 H$_2$O/IPA solution at 6$^{\circ}$C for 90~s followed by 10~s in IPA, the chip is loaded in a Plassys evaporator. A 2~min argon milling cleaning is implemented to ensure good electrical contact between the two metallic layers. Then the chip is evaporated with a 35~nm thick layer of Aluminium with an angle of -30$^{\circ}$, followed by 5~min of oxydation in 5~mbar of pure oxygen, and the evaporation of 100~nm of Aluminium with a +30$^{\circ}$ angle. After lift-off, the chip is baked at 200$^{\circ}$C for 1~h. The resulting junctions are of three types as summarized in table \ref{junctions}. The SQUID embedded in the transmon features a big junction in parallel with a tiny one, while the SNAIL embedded in the resonator features three big junctions in parallel with a small one (see Fig.~\ref{fig:fig1}).

\begin{figure}[]
\includegraphics[width=\linewidth,keepaspectratio]{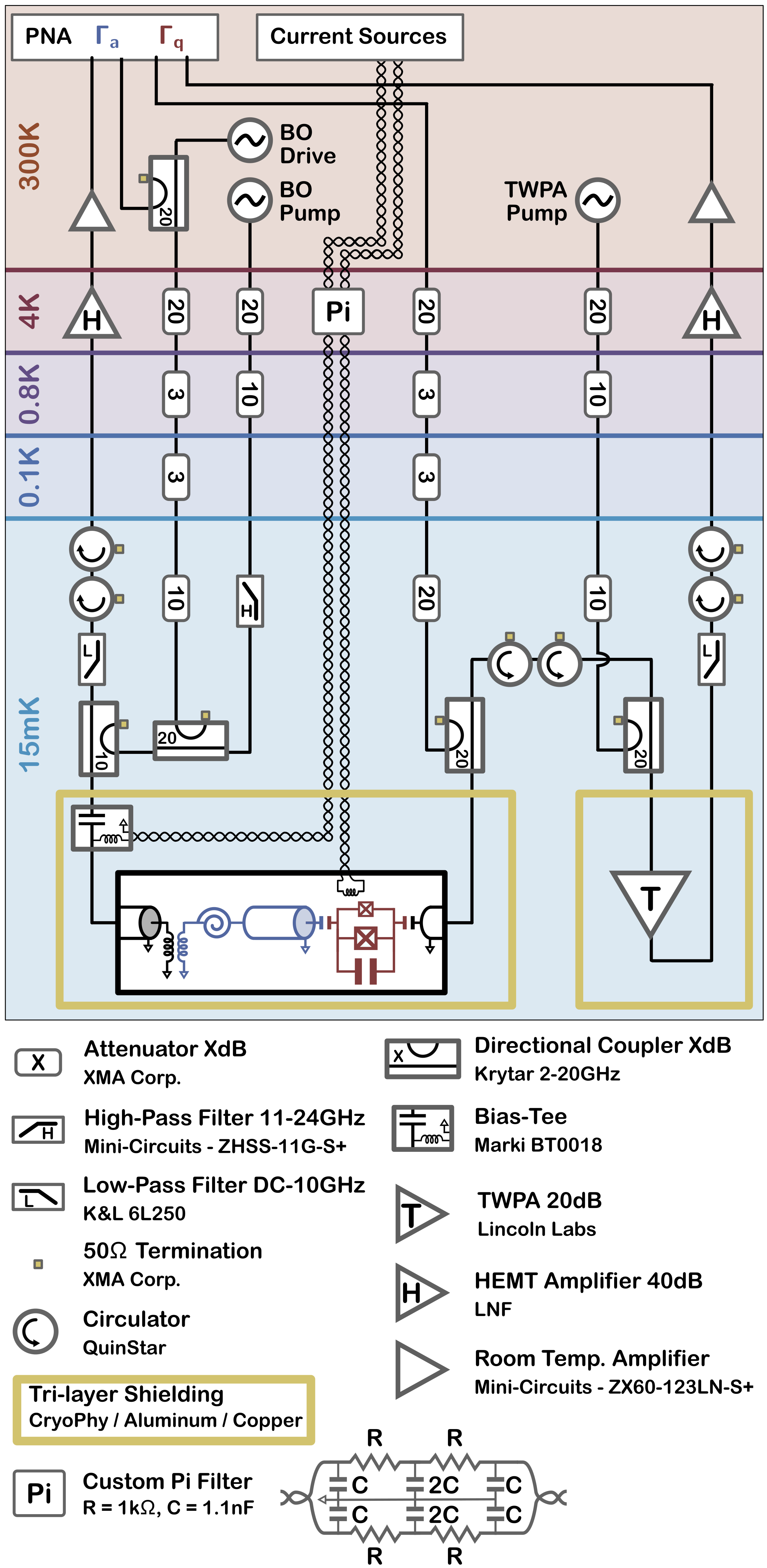}
\caption{Wiring diagram}
\label{wiring}
\end{figure}

\begin{table}[h]
\begin{tabular}{l c c c }
    \toprule
    \textbf{Junction type} & Big & Small & Tiny \\ 
    \toprule
    \textbf{Surface} [$\mu$m$^2$] & 2.10 & 0.14 & 0.08  \\
    \midrule
    \textbf{Inductance} [nH] & 0.19 & 2.57 & 4.48\\
    \bottomrule
\end{tabular}
\caption{Characteristics of the three junction types, as measured on test structures fabricated on the same chip.}
\label{junctions}
\end{table}

\subsection{Wiring}
\label{sec:wiring}
The sample is mounted in a microwave sample holder which was designed in-house, and tested to be free of spurious electromagnetic modes up to 15~GHz \cite{Villiers2023}. It is then mounted at the base plate of a Bluefors LD250 and cooled down to 15~mK. The wiring diagram is detailed on Figure \ref{wiring}. We use the four channels of a Keysight PNA N5222A to measure the reflection spectra of the resonator and transmon ports, denoted $\Gamma_\mathrm{a}$ and $\Gamma_\mathrm{q}$ respectively. Two DC current sources Yokogawa GS200 are used to bias the flux loops of the SNAIL and the SQUID with fluxes $\Phi_\mathrm{a}$ and $\Phi_q$. The Traveling Wave Parametric Amplifier (TWPA) provided by the group of W. Oliver at Lincoln Labs is powered by a R\&S SGS100A. It amplifies the transmon signal by about 20~dB away from its dispersive feature at 6.0~GHz. The tone that pumps the SNAIL is provided by an Agilent Technologies E8257D and travels to the sample either through the resonator PNA drive line, or through a distinct one. In order to maximize the amount of pump power reaching the sample around the parametric resonance $\omega_\mathrm{p}/2\pi\approx14$~GHz, without giving up on line attenuation at the resonance frequency $\omega_\mathrm{a}/2\pi\approx7$~GHz, we designed a dedicated microwave line for the pump. It includes a smaller amount of flat attenuation than the drive line, but features a high-rejection high-pass filter, with a pass-band from $11$~GHz to $24$~GHz. This pump line displays around 26~dB less attenuation than the drive line above 11~GHz, while maintaining sufficient attenuation around the oscillator frequency. These two orders of magnitude were crucial to approach instability in the BO regime $\lambda\sim|\delta_\mathrm{a}|\gg\kappa/2$, without heating up the cryostat. Finally a R\&S SMB100A provides the coherent drive on the resonator injecting photons to calibrate the dispersive interaction strength. All instruments are referenced to a Stanford Research Systems FS725 Rubidium clock.

\FloatBarrier
\section{Theory of the dispersive interaction of a qubit and squeezed photons}
\label{sec:theory}
In this appendix, we derive the pieces of theory relevant to the understanding of the interaction of a qubit with squeezed photons. To begin with, we will show how a SNAIL-resonator can emulate a Bogoliubov oscillator (BO) which hosts squeezed photons as eigenstates, and we will follow the procedure of Ref.~\cite{Steck2007} to derive the master equation and input-output relation for such excitations. Then we will derive a perturbative Hamiltonian capturing the interaction between a qubit and squeezed photons. Finally we will compute the relevant observables to the description of the spectral properties of a qubit interacting with an oscillator filled with squeezed photons, whether it is driven or not. Extension of these results to a transmon are presented in Appendix~\ref{sec:theory_trm}.

\subsection{Squeezed photons in a SNAIL-resonator}
\label{sec:theory_dpo}
First we consider a SNAIL-resonator biased at its Kerr-free flux point and strongly driven, or "pumped", close to the parametric resonance. At this specific operating point, it can be minimally described by an anharmonic oscillator with bare frequency $\omega_\mathrm{a}$ dressed by a third-order nonlinearity $g_3$, such that its Hamiltonian writes \cite{Sivak2019}:
\begin{equation}
%\begin{aligned}
\bm{\mathcal{H}}_\textbf{a}/\hbar = \omega_\mathrm{a} \bad \ba + g_3\big(\ba+\bad\big)^3
    - \varepsilon_\mathrm{p} \cos\omega_\mathrm{p} t \big(\ba + \bad \big)\;, 
%\end{aligned}
\label{Ha}
\end{equation}
%The pump amplitude is such that: $\varepsilon_\mathrm{p}=\mu\sqrt{P_p}$, with $P_p$ the applied pump power at room temperature and $\mu$ an unknown coefficient related to the attenuation of the microwave lines. 
 where $g_3\ll\omega_\mathrm{a}$. As customary for driven systems, we displace operator $\ba$ by its mean value $\ba\rightarrow \ba+\xi(t)$, where $\xi$ is a complex time-dependent parameter verifying $\dot{\xi}=-i\omega_\mathrm{a}\xi  -(\kappa/2)\xi +i\epsilon_p\cos(\omega_\mathrm{p}t)$ \cite{Boutin2017}. At times $t\gg 1/\kappa$, and in the regime where $\kappa\ll |\omega_\mathrm{a} \pm \omega_\mathrm{p}|$:
\begin{equation}
    \xi(t)\approx\frac{\varepsilon_\mathrm{p}/2}{\omega_\mathrm{a}-\omega_\mathrm{p}}e^{-i\omega_\mathrm{p}t} + \frac{\varepsilon_\mathrm{p}/2}{\omega_\mathrm{a}+\omega_\mathrm{p}}e^{i\omega_\mathrm{p}t}\;.
\end{equation}
Further going to a frame rotating at half the pump frequency through the unitary $\bm{\mathcal{U}_\omega}=\exp\{i\omega_\mathrm{p}t\bad\ba/2\}$, resulting in $\ba\rightarrow \ba e^{-i\omega_\mathrm{p}t/2}$, the transformed Hamiltonian exactly writes:
\begin{equation}
\begin{aligned}
\bm{\mathcal{H}^{\xi,\omega}}_\textbf{a}/\hbar = \delta_\mathrm{a} \bad \ba + g_3\Big(&\ba e^{-i\omega_\mathrm{p}t/2}+\bad e^{i\omega_\mathrm{p}t/2} \\
    &- \Pi e^{-i\omega_\mathrm{p}t} - \Pi^\ast e^{i\omega_\mathrm{p}t} \Big)^3\;,
\end{aligned}
\label{Hadisplaced}
\end{equation}
where $\delta_\mathrm{a}=\omega_\mathrm{a}-\omega_\mathrm{p}/2$. We place ourselves in the regime where $|\delta_\mathrm{a}|\ll \omega_\mathrm{a}$, hence $\Pi \approx \varepsilon_\mathrm{p}/3\omega_\mathrm{a}$. We next perform the rotating wave approximation (RWA), and define the time-averaged photon Hamiltonian as $\bm{\mathcal{H}}_\textbf{ph} \equiv \overline{\bm{\mathcal{H}^{\xi,\omega}}_\textbf{a}}$ \cite{Mirrahimi2015}. We find:
\begin{equation}
	\bm{\mathcal{H}}_\textbf{ph}/\hbar = \delta_\mathrm{a} \bad \ba - \frac{\lambda}{2} \left(\ba^2+\badd\right)\;,
\label{Harwa}
\end{equation}
where $\lambda\approx 2g_3\varepsilon_\mathrm{p}/\omega_\mathrm{a}$ is the two-photon pump amplitude. Thus a SNAIL-resonator pumped near the parametric resonance $2\omega_\mathrm{a}$ emulates a degenerate parametric oscillator (DPO) \cite{Carmichael1984}. The validity of the RWA is granted by $g_3 \Pi\ll\omega_\mathrm{p}$. In the case where $\lambda<|\delta_\mathrm{a}|$, the latter Hamiltonian can be diagonalized by means of a Bogoliubov transformation using the canonical basis $\bal = \ba\cosh r - \bad\sinh r$ where $r$ is the squeezing parameter defined by $\tanh2r=\lambda/|\delta_\mathrm{a}|$. This approach is equivalent to transforming the Hamiltonian through the squeezing unitary $\bm{\mathcal{U}_s}=e^{r/2(\ba^2-\ba^{\dagger 2})}$ by noting that $\bal=\bm{\mathcal{U}_s}^\dagger \ba \bm{\mathcal{U}_s}$. In this new basis the Hamiltonian \eqref{Harwa} writes:
\begin{equation}
    \bm{\mathcal{H}}_\textbf{ph}/\hbar = \Omega_\mathrm{a}[r]\bald \bal\;,
\end{equation}
where $\Omega_\mathrm{a}[r]=\delta_\mathrm{a}/\cosh2r$. One can also show that $\Omega_\mathrm{a}[r]=\sqrt{\delta_\mathrm{a}^2-\lambda^2}$. When $\lambda=0$, the Hamiltonian $\bm{\mathcal{H}}_\textbf{ph}$ is that of a simple harmonic oscillator, and its eigenstates are Fock states $\{\ket{n_\mathrm{a}}\}$ with eigenenergies $n_\mathrm{a}\delta_\mathrm{a}$, where $n_a$ are integers. Instead, when the two-photon pump is applied, the eigenstates are squeezed Fock states $\{\bm{\mathcal{U}_s}^\dagger \ket{n_\mathrm{a}}\}$ with eigenenergies $n_a\Omega_\mathrm{a}[r]$ (see Fig.~\ref{fig:fig1}). %The latter description is resilient to strong two-photon pump amplitudes $\lambda\gg\kappa/2$, provided $\lambda<|\delta_\mathrm{a}|$. In this regime of parameters the DPO turns into a Bogoliubov oscillator (BO), whose eigenstates are squeezed photons.

%\begin{figure}
%\caption{Wigners?}
%\end{figure}
%\FloatBarrier

\subsection{Input-output theory for squeezed photons}
\label{sec:theory_inout}
The resonator drive is applied through a coupled feedline hosting a continuum of modes $\{\bm{c(\omega)}\}_\omega$ which will ultimately interact with mode $\ba$. It can be thought of as set of harmonic oscillators at all possible frequencies $\omega\in[0,+\infty)$ described by the Hamiltonian: $\bm{\mathcal{H}}_\textbf{bath} = \int\mathrm{d}\omega' \hbar\omega' \bm{c(\omega')}^\dagger \bm{c(\omega')}$, where $\left[\bm{c(\omega)}, \bm{c(\omega')}^\dagger\right]=\bm{\delta(\omega-\omega')}$. In order to account for the evolution of the SNAIL-resonator opened to its environment, the dynamics of the total system $\{\text{resonator}+\text{bath}\}$ needs to be addressed through $\bm{\mathcal{H}}_\textbf{tot} = \bm{\mathcal{H}}_\textbf{a} + \bm{\mathcal{H}}_\textbf{bath} + \bm{\mathcal{H}}_\textbf{int}$ where:
\begin{equation}
    \bm{\mathcal{H}}_\textbf{int}/\hbar = \sqrt{\frac{\kappa}{2\pi}} \int \mathrm{d}\omega' \big(\ba+\bad\big) \big[ \bm{c(\omega')}  + \bm{c(\omega')}^\dagger \big],
    \label{Hint} 
\end{equation}
such that each mode $\bm{c(\omega)}$ is coupled to the cavity at a rate $\kappa$. The latter expression assumes the Markov approximation which neglects the frequency dependence of the coupling constant: $\kappa(\omega)\approx\kappa$. This approximation is well verified in our experiment since the impedance of the transmission line is almost flat over the frequency window (of order $\kappa$) sampled by the resonator.

We follow the same treatment for the total Hamiltonian as in the previous subsection. While the bath part is trivially modified, the transformed interaction part writes in the Bogoliubov basis:
\begin{align}
    \overline{\bm{\mathcal{H}^{\xi,\omega}}_\textbf{int}}/\hbar = &\sqrt{\frac{\kappa}{2\pi}}\int\mathrm{d}\omega' \Big\{ \cosh r \left[\bal \bm{c(\omega')}^\dagger + \bald\bm{c(\omega')} \right] \nonumber \\
    &+ \sinh r \left[\bal \bm{c(\omega')} + \bald\bm{c(\omega')}^\dagger \right] \Big\}\;.
    \label{eq:Hint_alpha}
\end{align}
At this stage the RWA is valid as long as: $\kappa e^r/2 \ll \omega_\mathrm{p}$, a regime safely maintained for all squeezing values. Just like the BO displayed enhanced interactions to a target qubit mode (Eq.~\ref{eq:HRabi}), its enhanced interactions to the bath modes are immediately apparent in Eq.~\eqref{eq:Hint_alpha}. Then we can write the equations of motion for the Heisenberg operators $\bm{\alpha(t)}$ and $\bm{c(\omega')(t)}$:
\begin{subequations}
\begin{align}
	\partial_t \bm{\alpha(t)} =& -i\Omega_\mathrm{a}[r] \bm{\alpha(t)} \label{g_langevin} \\ &- i\sqrt{\frac{\kappa}{2\pi}} \int \mathrm{d}\omega' \bm{c(\omega')} \cosh r + \bm{c(\omega')}^\dagger \sinh r\;,\nonumber \\
	\partial_t \bm{c(\omega')} =& -i \omega'\bm{c(\omega')} \label{c_langevin} \\ &- i \sqrt{\frac{\kappa}{2\pi}} \left( \bm{\alpha(t)} \cosh r + \bm{\alpha(t)}^\dagger \sinh r \right)\;, \nonumber
\end{align}
\end{subequations}
where the explicit time-dependence of the operator $\bm{c(\omega)(t)}$ has been omitted. Integrating equation \eqref{c_langevin} from a past reference time $t_0$ until the experiment time $t$, and defining the input field operator as $\textbf{a}_\mathrm{in}\bm{(t)} = (-i/\sqrt{2\pi}) \int \mathrm{d}\omega' \bm{c(\omega')(t_0)} e^{-i\omega'(t-t_0)}$, we can rewrite equation \eqref{g_langevin} as:
\begin{equation}
\begin{aligned}
	\partial_t \bm{\alpha(t)} = &-i\Omega_\mathrm{a}[r] \bm{\alpha(t)} - \frac{\kappa}{2}\bm{\alpha(t)} \\ &+ \sqrt{\kappa} \left(\textbf{a}_\mathrm{in}\bm{(t)} \cosh r  - \textbf{a}_\mathrm{in}\bm{(t)}^\dagger \sinh r  \right) .
	\label{QLE}
 \end{aligned}
\end{equation}
Equation \eqref{c_langevin} could also have been integrated from a future time $t_1$ until the experiment time $t$, defining the output filed operator: $\textbf{a}_\mathrm{out}\bm{(t)} = (i/\sqrt{2\pi}) \int \mathrm{d}\omega' \bm{c(\omega')(t_1)} e^{-i\omega'(t-t_1)}$. The input and output fields satisfy the closure relation:
\begin{equation}
	\textbf{a}_\mathrm{out}\bm{(t)} + \textbf{a}_\mathrm{in}\bm{(t)} = \sqrt{\kappa} \big( \bm{\alpha(t)} \cosh r + \bm{\alpha(t)}^\dagger \sinh r \big) .
	\label{inputout}
\end{equation} 
The input and output fields have zero mean, and obey the commutation relations: $\left[\textbf{a}_\mathrm{in}\bm{(t)}, \textbf{a}_\mathrm{in}\bm{(t')}^\dagger\right] = \bm{\delta(t-t')}$, $\left[\textbf{a}_\mathrm{in}\bm{(t)}, \textbf{a}_\mathrm{in}\bm{(t')}\right] = 0$ (same for $\textbf{a}_\mathrm{out}\bm{(t)}$). The temperature of the environment is defined through the thermal occupancy $\bar{n}_\mathrm{th}$ such that $\braket{\textbf{a}_\mathrm{in}\bm{(t)}^\dagger \textbf{a}_\mathrm{in}\bm{(t')}} = \bar{n}_\mathrm{th}\bm{\delta(t-t')}$ and $\braket{\textbf{a}_\mathrm{in}\bm{(t)} \textbf{a}_\mathrm{in}\bm{(t')}^\dagger} = \left(1+\bar{n}_\mathrm{th}\right) \bm{\delta(t-t')}$. In the case where the oscillator is driven with a coherent tone of amplitude $\varepsilon_d$ at a frequency $\omega_d$, the input operator needs to be displaced by a classical contribution: $\textbf{a}_\mathrm{in}\bm{(t)} - i(\varepsilon_d^\ast/2\sqrt{\kappa})e^{-i(\omega_d-\omega_\mathrm{p}/2)t}$.

Together, the quantum Langevin equation \eqref{QLE} and the input-output relation \eqref{inputout} fully capture the dynamics of the squeezed photons in contact with their environment. It is here described in terms of the incoming and outcoming fields of the bare mode $\ba$, which correspond to the physical port used to drive and read-out the BO. Interestingly the decay rate of these squeezed photons does not change with squeezing, a property that can be traced back to their enhanced interactions to the bare bath modes (Eq.~\ref{eq:Hint_alpha}). In turn, a BO operated as an amplifier is not constrained by a constant gain-bandwidth product (see Fig.~\ref{fig:SPA_gbw}) \cite{Villiers2023}.

\subsection{Dispersive transformation}
\label{sec:theory_qubit}
The coupling of the BO with a qubit is now addressed. Following the main text, the qubit with frequency $\omega_\mathrm{q}$ is introduced through the Pauli operators $(\sgz, \sgm, \sgp)$. In a rotating frame at frequency $\omega_\mathrm{p}/2$ for both modes, and assuming the BO-qubit coupling to be small ($g \ll \omega_\mathrm{a}, \omega_\mathrm{q}$), the system can be described by a Jaynes-Cummings Hamiltonian augmented by a squeezing term:
\begin{equation}
\begin{aligned}
    \bm{\mathcal{H}}_\textbf{q-ph}/\hbar =\:& \delta_\mathrm{a} \ba^\dagger\ba - \frac{\lambda}{2}\left( \ba^2 + \badd \right) + \delta_\mathrm{q} \frac{\sgz}{2} \\
    &+ g \left( \ba \sgp + \bad\sgm \right)\;,
\end{aligned}
\label{eq:H_JC}
\end{equation}
where $\delta_\mathrm{q}=\omega_\mathrm{q}-\omega_\mathrm{p}/2$. Continuing with a diagonalization of the oscillator-only part of the Hamiltonian, we find in the Bogoliubov basis:
\begin{align}
    \bm{\mathcal{H}}&_\textbf{q-ph}/\hbar = \Omega_\mathrm{a}[r] \bald \bal + \delta_\mathrm{q} \frac{\sgz}{2} \\
    + g&\cosh r \left( \bal \sgp + \bald\sgm \right) + g\sinh r \left( \bal \sgm + \bald\sgp \right)\;. \nonumber
\end{align}
While Ref.~\cite{Qin2018, Leroux2018} focused on the resonant limit $\delta_\mathrm{q}\approx\Omega_\mathrm{a}[r]$, we place ourselves in the dispersive regime. Owing to the presence of the pump mixing signal and idler photons, the dispersive interaction to a BO is restricted to the regime where:
\begin{subequations}
\begin{align}
        \Delta[r] &=\delta_\mathrm{q} - \Omega[r] \gg g e^r \;, \\
        \Sigma[r] &=\delta_\mathrm{q} + \Omega[r] \gg g e^r \;.
\end{align}
\end{subequations}
Not only the qubit needs to be far from resonance with the BO signal frequency, but also with the mirror idler one. There, we use a Schrieffer-Wolff (SW) transformation to write the Hamiltonian in a basis that decouples the qubit and the Bogoliubov mode, at first order in coupling over the detunings. The generator of this transformation writes:
\begin{equation}
    %\bm{\mathcal{S}} = \frac{g\cosh r}{\Delta[r]} \left(\bg \sgp - \bgd\sgm \right) - \frac{g\sinh r}{\Sigma[r]} \left(\bg\sgm- \bgd\sgp \right),
    \bm{\mathcal{S}} = \frac{g\cosh r}{\Delta[r]} \bal \sgp - \frac{g\sinh r}{\Sigma[r]} \bal\sgm - \text{h.c.}
\end{equation}
In the transformed basis $\bal \rightarrow e^{-\bm{\mathcal{S}}} \bal e^{\bm{\mathcal{S}}}$, $\sgz \rightarrow e^{-\bm{\mathcal{S}}} \sgz e^{\bm{\mathcal{S}}}$, the Hamiltonian writes at second order in $g/\Delta[r]$, $g/\Sigma[r]$:
\begin{align}
    \bm{\mathcal{H}}_\textbf{q-ph}/\hbar &= \Omega_\mathrm{a}[r] \bald\bal + \delta_\mathrm{q} \frac{\sgz}{2} \label{Hchia} \\
    + \chi[r]& \left(\bald\bal + \frac{1}{2}\right) \frac{\sgz}{2}
    + \chi_a[r] \left(\bal^2 + \baldd \right) \frac{\sgz}{2}\;, \nonumber
\end{align}
where the dispersive interaction strength $\chi[r]$ and its anomalous counterpart $\chi_a[r]$ write:
\begin{subequations}
\begin{align}
    \chi[r] &= \frac{2g^2\cosh^2r}{\delta_\mathrm{q}-\Omega_\mathrm{a}[r]} + \frac{2g^2\sinh^2r}{\delta_\mathrm{q}+\Omega_\mathrm{a}[r]} \;, \\
    \chi_a[r] &= \frac{g^2\sinh2r}{\delta_\mathrm{q}}\frac{\delta_\mathrm{q}^2}{\delta_\mathrm{q}^2-\Omega_\mathrm{a}^2[r]}\;.
\end{align}
\end{subequations}
Further assuming $\chi_a[r]\ll|2\Omega_\mathrm{a}[r]|$, Hamiltonian~\eqref{Hchia} can be approximated by its secular part, which yields Eq.~\eqref{eq:Hchi} of the main text.
%\begin{equation}
%    \bm{\mathcal{H}}_\textbf{q-ph}/\hbar = \Omega_\mathrm{a}[r]\bald \bal 
%    + \left[\delta_\mathrm{q} + \chi[r] \Big(\bald \bal + \frac{1}{2} \right) \Big] \frac{\sgz}{2}\;.
%\label{Hchi}
%\end{equation}
%This approximation neglects the mixing of the signal and idler peaks induced by the qubit, an approximation even more valid as the BO is far from coalescence.

Similarly the loss operators are dressed by the SW transformation. Introducing the dimensionless parameter $\eta$ such that: $ge^r, \: \kappa, \: \Gamma_1, \: \Gamma_\phi < \eta \times \mathrm{min} ( |\Delta[r]|, \: |\Sigma[r]|)$, and under the assumption that $\eta \ll 1$, these composite loss operators can be split into independant channels up to second order in $\eta$:
\begin{subequations}
\begin{align}
    \bm{L}_\textbf{a} &= \sqrt{\kappa}\left(\cosh r \bal + \sinh r \bald \right) \;, \label{kappa}\\
    \bm{L^-}_\textbf{a} &= \sqrt{\kappa}\left(\frac{g\cosh^2r}{\Delta[r]} - \frac{g\sinh^2r}{\Sigma[r]}\right)\sgm \;, \label{purcellm} \\
    \bm{L^+}_\textbf{a} &= \sqrt{\kappa} g\cosh r \sinh r \left(\frac{1}{\Delta[r]} - \frac{1}{\Sigma[r]}\right)\sgp\;, \label{purcellp} \\
    \bm{L_-} &= \sqrt{\Gamma_1} \sgm \;, \label{gamma1} \\
    \bm{L_-^\phi} &= \sqrt{\Gamma_1} \left(\frac{g\cosh r}{\Delta[r]} \bal + \frac{g\sinh r}{\Sigma[r]} \bald \right)\sgz\;, \label{rel_deph} \\
    \bm{L_\phi} &= \sqrt{\frac{\Gamma_\phi}{2}} \sgz \;, \label{gammaphi} \\
    \bm{L_\phi^+} &= \sqrt{\frac{\Gamma_\phi}{2}} \left(\frac{2g\cosh r}{\Delta[r]}\bal + \frac{2g\sinh r}{\Sigma[r]}\bald \right)\sgp \;, \label{rel_ex}\\
    \bm{L_\phi^-} &= \sqrt{\frac{\Gamma_\phi}{2}} \left(\frac{2g\cosh r}{\Delta[r]}\bald + \frac{2g\sinh r}{\Sigma[r]}\bal \right)\sgm \;. \label{rel_rel}
\end{align}
\end{subequations}
Among these loss operators we identify the Purcell relaxation of the qubit through the Bogoliubov mode \eqref{purcellm}, and its induced excitation counterpart \eqref{purcellp}, along with the cavity dressed dephasing \eqref{rel_deph}, excitation \eqref{rel_ex} and relaxation \eqref{rel_rel} \cite{Boissonneault2009}. When entering a Lindblad master equation, the amplitudes of the aforementioned processes are of order 3 in $\eta$. Hence at second order, only the bare loss operators \eqref{kappa}, \eqref{gamma1}, \eqref{gammaphi} need to be considered.

Finally, it is instructive to look at the dispersive interaction strength in the limit $|\delta_\mathrm{q}|\gg|\delta_\mathrm{a}|$. In this regime, the renormalization of the oscillator frequency is negligible when compared to the qubit-oscillator detuning, such that $\Delta[r]\approx\Sigma[r]\approx\delta_\mathrm{q}$. Denoting the bare interaction parameter by $\chi_\infty=2g^2/\delta_\mathrm{q}$, the enhanced dispersive interaction strength reads $\chi[r]\approx\chi_\infty\cosh2r$.
%Hence in the BO regime, the coupling rate to the Bogoliubov mode follows closely the instability of the oscillator.

\subsection{Spectroscopy of a qubit interacting with squeezed photons}
\label{sec:theory_mid}
We consider a BO continuously squeezed, and coherently driven at its renormalized frequency. At long times, the occupation of the BO converges towards a mean-photon number $\bar{n}_\alpha$, and fluctuates by $\delta n_\alpha(t)$. The statistical properties of the BO occupancy reflect both the effects of the squeezing and the coherent drive. Computing the impact of this mixed statistics on a dispersively coupled qubit is the topic of this part \cite{Blais2004, Gambetta2006, Lemonde2016}.

A qubit initialized in a coherent superposition of its basis states at a time $t_0$ will pickup a relative phase according to its dispersive interaction with the BO (see Eq.~\eqref{eq:Hchi}). After an interaction time $t$, we write this phase: $\varphi(t) \equiv \bar{\varphi} + \delta\varphi(t)$. The mean part reads:
\begin{equation}
    \bar{\varphi} = \left(\delta_\mathrm{q} + \frac{\chi[r]}{2} + \chi[r]\bar{n}_\alpha\right)(t-t_0)\;,
\end{equation}
which displays the Lamb-shifted qubit detuning $\delta_\mathrm{q}+\chi[r]/2$, and the AC-Stark contribution $\chi[r]\bar{n}_\alpha$. The fluctuating part reads:
\begin{equation}
    \delta\varphi(t) = \chi\int_{t_0}^{t_0+t}\mathrm{d}\tau\delta n_\alpha(\tau)\;,
\end{equation}
and its randomness is at the heart of the dephasing mechanism. As the BO excitations are short-lived compared to the typical qubit-BO interaction time ($\kappa\gg|\chi[r]|$), $\delta\varphi$ can be thought of as a sum of independent random variables. Hence the central limit theorem applies, and $\delta\varphi$ follows a Gaussian distribution. Since $\delta n_\alpha$ has zero mean, so does $\delta\varphi$. The induced dephasing by the BO on the qubit $\Delta\gamma_\phi$ is commonly defined as $e^{-\Delta\gamma_\phi t} \equiv \braket{\braket{e^{i\delta\varphi}}}$, where $\braket{\braket{\boldsymbol{ \cdot}}}$ refers to the average over multiple noise realizations (statistical ensemble average). Owing to the previously detailed statistics of $\delta\varphi$, we find that $\braket{\braket{e^{i\delta\varphi}}} = e^{-\frac{1}{2}\braket{\braket{\delta\varphi^2}}}$, so that the induced dephasing reads:
\begin{align}
        &\Delta\gamma_\phi = \frac{\chi^2[r]}{2t} \iint_{t_0}^{t_0+t}\mathrm{d}t_1\mathrm{d}t_2 \braket{\delta n_\alpha(t_2) \delta n_\alpha(t_1)}\;, \label{eq:induced} \\
        &\braket{\delta n_\alpha(t_2) \delta n_\alpha(t_1)}=\braket{(\bm{n_\alpha(t_2)}-\bar{n}_\alpha)(\bm{n_\alpha(t_1)}-\bar{n}_\alpha)} \;, \nonumber
\end{align}
where $\bm{n_\alpha(t)}=\bm{\alpha^\dagger(t)\alpha(t)}$. To lowest order in $\chi/\kappa$, the average $\braket{\bm{\cdot}}$ denotes the expectation value of the uncoupled system. Elucidating the dispersive and dissipative effects of the BO on the qubit amounts to solving the quantum Langevin equation \eqref{QLE}, and computing the mean-photon number at long times $\bar{n}_\alpha=\braket{\bald\bal}$, and the correlation function $\mathcal{C}(t_1, t_2) = \braket{\delta n_\alpha(t_2) \delta n_\alpha(t_1)} = \braket{\bm{n_\alpha(t_2)}\bm{n_\alpha(t_1)}} -\bar{n}_\alpha^2$.

In the presence of a coherent drive of amplitude $\epsilon_d$ at the BO resonance, Eq.~\eqref{QLE} is most conveniently solved in a displaced frame. Specifically we write $\bm{\alpha(t)}=\ual(t) + \bm{d(t)}$, where $\ual(t)$ solves the classical part of Eq.~\eqref{QLE}, and $\bm{d(t)}$ its quantum part. The displaced Bogoliubov operator $\bd$ follows the same commutation relations as the original one. We find:
\begin{widetext}
\begin{subequations}
\begin{align}
    \ual(t) &= \ual(t_0)e^{-(i\Omega_\mathrm{a}+\kappa/2)(t-t_0)} - i \int_{t_0}^t \mathrm{d}\tau \left\{\frac{\varepsilon^\ast_d}{2}e^{-i\Omega_\mathrm{a}\tau}\cosh r + \frac{\varepsilon_d}{2} e^{i\Omega_\mathrm{a}\tau}\sinh r \right\}e^{-(i\Omega_\mathrm{a}+\kappa/2)(t-\tau)} \;,\\
    \bm{d(t)} &= \bm{d(t_0)}e^{-(i\Omega_\mathrm{a}+\kappa/2)(t-t_0)} + \sqrt{\kappa}\int_{t_0}^t\mathrm{d}\tau \Big\{\bm{a_\mathrm{in}(\tau)}\cosh r - \bm{a_\mathrm{in}^\dagger(\tau)}\sinh r \Big\}e^{-(i\Omega_\mathrm{a}+\kappa/2)(t-\tau)}\;.
\end{align}
\end{subequations}
\end{widetext}
Thus the mean-photon number can be readily computed as $\bar{n}_\alpha = |\ual(t)|^2 + \braket{\bm{d^\dagger(t)}\bm{d(t)}}$. Moreover, owing to the quadratic nature of the system-bath Hamiltonian, we can use Wick's theorem to compute the correlation function:
\begin{align}
    \mathcal{C}(t_1, t_2) 
    =&\; \braket{\bm{d_2}^\dagger\bm{d_1}^\dagger} \braket{\bm{d_2}\bm{d_1}}
    + \braket{\bm{d_2}^\dagger\bm{d_1}} \braket{\bm{d_2}\bm{d_1}^\dagger} \nonumber \\
    &+ \ual_2^\ast \ual_1^\ast \braket{\bm{d_2}\bm{d_1}} + \ual_2 \ual_1 \braket{\bm{d_2}^\dagger\bm{d_1}^\dagger} \\
    &+ \ual_2 \ual_1^\ast \braket{\bm{d_2}^\dagger\bm{d_1}} + \ual_2^\ast \ual_1 \braket{\bm{d_2}\bm{d_1}^\dagger} \;, \nonumber
\end{align}
where $\bm{d_i}=\bm{d(t_i)}$ and $\ual_i=\ual(t_i)$. We then focus at the long time limit $t\ge t_0\gg\kappa^{-1}$ for which the BO converges to a limit cycle with amplitude: 
\begin{equation}
\ual(t) = -i\frac{\varepsilon_d^\ast\cosh r}{\kappa}e^{-i\Omega_\mathrm{a}t} -i\frac{\varepsilon_d\sinh r}{\kappa+4i\Omega_\mathrm{a}}e^{i\Omega_\mathrm{a}t}\;.
\end{equation}
In the limit $|\Omega_\mathrm{a}[r]|\gg\kappa/2$ (far from coalescence), the classical part of the Bogoliubov mode reduces to an amplified coherent signal $\ual(t) \approx -ie^{-i\Omega_\mathrm{a}t}(\varepsilon_d^\ast/\kappa)\cosh r$. Indeed, in that regime, the BO induces negligible mixing between the signal and idler components of the drive. Next we turn to the statistical properties of the quantum part in the long time limit $t_1,t_2\ge t_0\gg\kappa^{-1}$:
\begin{widetext}
\begin{subequations}
    \begin{align}
        \braket{\bm{d^\dagger(t_2)}\bm{d(t_1)}} &= \left(\sinh^2r + \bar{n}_\mathrm{th} + 2\bar{n}_\mathrm{th}\sinh^2r\right)e^{i\Omega_\mathrm{a}(t_2-t_1)}e^{-\kappa|t_2-t_1|/2}\:, \\
        \braket{\bm{d(t_2)}\bm{d^\dagger(t_1)}} &= \left(1+\sinh^2r+ \bar{n}_\mathrm{th} + 2\bar{n}_\mathrm{th}\sinh^2r\right)e^{-i\Omega_\mathrm{a}(t_2-t_1)}e^{-\kappa|t_2-t_1|/2}\:, \\
        \braket{\bm{d(t_2)}\bm{d(t_1)}} &= i\frac{\kappa}{4\Omega_\mathrm{a}}\frac{\sinh2r}{1-i\kappa/2\Omega_\mathrm{a}}\left(1+2\bar{n}_\mathrm{th}\right)e^{-(i\Omega_\mathrm{a}+\kappa/2)|t_2-t_1|}\:. \label{eq:anomalous}
    \end{align}
\end{subequations}
\end{widetext}

First, we focus on the situation where the environment is held in vacuum: $\bar{n}_\mathrm{th}=0$. At zeroth order in $\eta=\kappa\sinh2r/4|\Omega_\mathrm{a}[r]|$, the anomalous correlator~\eqref{eq:anomalous} vanishes, and the displaced Bogoliubov mode $\bd$ resembles a thermal field with occupancy $\sinh^2r$ \cite{Lemonde2016}. At first order in $\eta$ we find $\bar{n}_\alpha \approx \bar{n}_\mathrm{d}\cosh^2 r + \sinh^2r$, where $\bar{n}_\mathrm{d}=|\varepsilon_d|^2/\kappa^2$ is the number of circulating photons that the coherent drive would maintain in the oscillator in the absence of squeezing. Far from coalescence, the mean occupation of the BO results from the sum of the amplified drive and the effective thermal population. Second, we look at the correlation function, either for a squeezed oscillator in contact with vacuum (up to first order in $\eta$), or for a regular oscillator in contact with a hot environment:
\begin{widetext}
\begin{subequations}
    \begin{align}
        \mathcal{C}(t_1,t_2|r,\bar{n}_\mathrm{th}=0) &\approx \sinh^2r \big(1+\sinh^2r\big)e^{-\kappa|t_2-t_1|} + \bar{n}_\mathrm{d}\cosh^2r \big(1 + 2\sinh^2r \big) e^{-\kappa|t_2-t_1|/2}\;, \label{eq:correl}\\
        \mathcal{C}(t_1,t_2|r=0,\bar{n}_\mathrm{th}) &= \bar{n}_\mathrm{th} \big(1+\bar{n}_\mathrm{th}\big)e^{-\kappa|t_2-t_1|} +  \bar{n}_\mathrm{d} \big( 1 + 2 \bar{n}_\mathrm{th}\big)e^{-\kappa|t_2-t_1|/2}\;.
    \end{align}
\end{subequations}
\end{widetext}
Note that the correlation function~\eqref{eq:correl} also features oscillatory terms proportional to $ \mathrm{Re}\big(\eta e^{2i\Omega_\mathrm{a}t_1-\kappa|t_2-t_1|/2}\big)$ that we omitted here, anticipating on the averaging performed when computing the induced dephasing. Comparing these two correlation functions lets us confirm the resemblance of a BO with a hot oscillator with thermal occupancy $\sinh^2r$. 

Finally, we can write the frequency shift of a qubit dispersively coupled to a driven BO far from coalescence as $\Delta\omega_\mathrm{q}=\Delta\omega_\mathrm{q}[r] + \Delta\omega_\mathrm{q}[r,\bar{n}_\mathrm{d}]$ where:
\begin{subequations}
\begin{align}
    \Delta\omega_\mathrm{q}[r] &= \chi[r]\Big(\frac{1}{2} + \sinh^2r\Big)\;, \\
    \Delta\omega_\mathrm{q}[r,\bar{n}_\mathrm{d}] &= \chi[r] \bar{n}_\mathrm{d}\cosh^2 r \;. \label{eq:ac_squeeze}
\end{align}
\label{eq:shift_qubit}
\end{subequations}
The first contribution amounts to a modified Lamb shift accounting for the equivalent thermal occupation of the BO. The second contribution is an AC-Stark shift accounting for the amplification of the input drive by the BO anti-squeezing. Similarly the induced dephasing reads $\Delta\gamma_\phi = \Delta\gamma_\phi[r] + \Delta\gamma_\phi[r,\bar{n}_\mathrm{d}]$ where:
\begin{subequations}
\begin{align}
    \Delta\gamma_\phi[r] &= \frac{\chi^2[r]}{\kappa} \sinh^2 r \left(1+\sinh^2r\right)\;, \\ 
    \Delta\gamma_\phi[r,\bar{n}_\mathrm{d}] &= \frac{2\chi^2[r]}{\kappa} \left(1 + 2\sinh^2r\right)\bar{n}_\mathrm{d} \cosh^2r\;. \label{eq:mid_squeeze}
\end{align}
\label{eq:broad_qubit}
\end{subequations}
We can map the first term to the characteristic dephasing of a qubit dispersively coupled to a hot oscillator \cite{Bertet2005, Rigetti2012}. The second term features the induced dephasing of a qubit measured by an amplified coherent drive on the oscillator, plus a cross term related to the equivalent BO thermal population. These equations are derived for a two-level system, and are adapted for a transmon in Appendix~\ref{sec:theory_trm}.

\FloatBarrier
\section{Degenerate parametric oscillator calibrations}
\label{sec:dpo_cal}
In this appendix, we present the calibration of the Kerr-free flux point of the SNAIL-resonator, necessary to operate it as a DPO. Then we turn to the description of its microwave response, and show how we can use it to calibrate the two-photon pump amplitude, whether the two-photon pump frequency matches the degenerate parametric resonance or not. Finally we discuss the various definitions of squeezing, whether it is enforced via a detuned pump or not.

\subsection{Kerr-free flux point of a SNAIL-resonator}
\label{sec:dpo_cal_kerr_free}
\begin{figure}[]
\includegraphics[width=\linewidth,keepaspectratio]{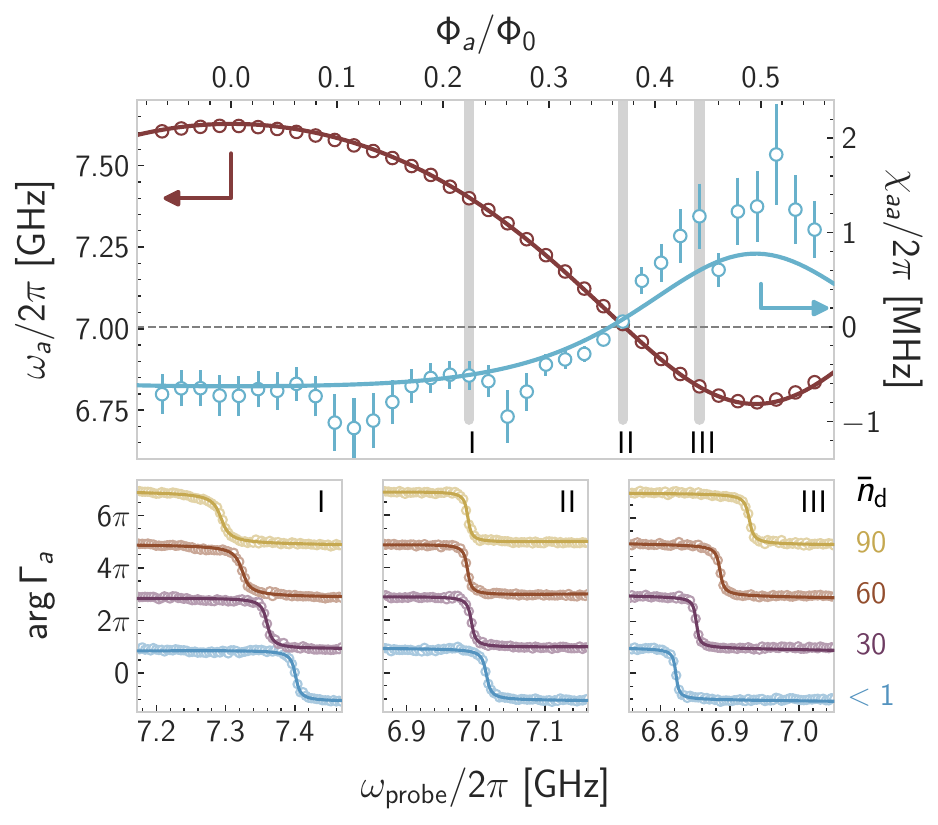}
\caption{Kerr spectroscopy of the oscillator. Top: measured resonant frequency (left axis) and Kerr nonlinearity (right axis) of the oscillator versus flux threaded through the SNAIL loop (x-axis), in units of flux quantum $\Phi_0=h/2e$. Fitting the data (open circles) to the theory extracted from Hamiltonian diagonalization  (full lines) sets all oscillator circuit parameters. Bottom: phase response (y-axis) of the oscillator to a weak probe of variable frequency (x-axis) when populated by a +300 MHz detuned drive with increasing power (bottom to top, curves offset for clarity) in units of circulating photon number $\bar{n}_\mathrm{d}$ indicated on the right. Each panel corresponds to a flux point where the Kerr non-linearity is negative (I), close to zero (II) and positive (III). Fitted response (full lines) are overlaid to the data (open circles).}
\label{fig:SPA_phi}
\end{figure}

Following \cite{Frattini2017}, a SNAIL-resonator is most generally described by the Hamiltonian:
\begin{equation}
    \bm{\mathcal{H}}_\textbf{a}/\hbar = \omega_\mathrm{a}(\Phi_\mathrm{a}) \ba^\dagger \ba + \sum_{m\ge 3} g_m(\Phi_\mathrm{a}) \left(\ba + \ba^\dagger \right)^m\;,
\end{equation}
where $g_m(\Phi_\mathrm{a})$ is the m$^{\text{th}}$-order nonlinearity inherited from the SNAIL potential energy, depending on the flux $\Phi_\mathrm{a}$ threading its loop. The fourth-order term of this expansion contributes to the Kerr nonlinearity of the oscillator. Owing to the specific choice of SNAIL parameters (see Table \ref{junctions}), the Kerr amplitude vanishes at a given flux point \cite{Frattini2018}. We identify this specific flux point by performing a Kerr spectroscopy of the oscillator (Fig.~\ref{fig:SPA_phi}). At each flux bias, we set a microwave drive 300 MHz above resonance populating the oscillator with increasing photon number $\bar{n}_\mathrm{d}$ (calibration in Appendix~\ref{sec:disp_cal}), and acquire its reflection spectrum. The resonance frequency shift $\Delta\omega_\mathrm{a}=\chi_\mathrm{aa}\bar{n}_\mathrm{d}$ is a direct measure of the Kerr amplitude $\chi_\mathrm{aa}$. As depicted in the insets of Fig.~\ref{fig:SPA_phi}, its value can be positive, negative, and set close to zero. In the dataset of Fig.~\ref{fig:SPA_phi} we find a Kerr-free point at $\omega_\mathrm{a}^0/2\pi=7.015$~GHz. Later in the cooldown, this operation point drifted to $\omega_\mathrm{a}^0/2\pi=6.940$~GHz which is used in the rest of the paper. The Kerr and resonance frequency versus flux pin down the non-linear resonator circuit parameters. From these parameters, we estimate a three-wave mixing amplitude at the Kerr-free point  $g_3/2\pi=18$~MHz. 

\subsection{Microwave response and two-photon pump calibration}
Following Appendix~\ref{sec:theory_inout}, the QLE for the bare oscillator Heisenberg operator $\textbf{a}\bm{(t)}$ writes at the Kerr-free flux point:
\begin{equation}
    \partial_t\textbf{a}\bm{(t)} = \frac{i}{\hbar}\left[\bm{\mathcal{H}}_\textbf{ph}, \textbf{a}\bm{(t)}\right] - \frac{\kappa}{2}\textbf{a}\bm{(t)} + \sqrt{\kappa}\textbf{a}_\textbf{in}\bm{(t)}\;,
\label{QLE_a}
\end{equation}
where $\bm{\mathcal{H}}_\textbf{ph}$ was defined in Eq.~\eqref{Harwa}, and $\kappa$ is the coupling rate of the oscillator to its feedline. Since the oscillator is overcoupled to its feedline, no other dissipation channel needs to be included. As we measure in reflection, the input-output relation reads: $\textbf{a}_\textbf{out}\bm{(t)}+\textbf{a}_\textbf{in}\bm{(t)} = \sqrt{\kappa}\textbf{a}\bm{(t)}$. We compute the complex output and express in the following form:  $\textbf{a}_\textbf{out}\bm{[\omega]}=\Gamma_\mathrm{a}[\omega]\textbf{a}_\textbf{in}\bm{[\omega]} + \Gamma_i[\omega]\textbf{a}_\textbf{in}\bm{[\text{-}\omega]}^\dag $, where $\Gamma_\mathrm{a}[\omega]$ is the complex signal gain response, and $\Gamma_i[\omega]$ is the complex idler gain response. We find:
\begin{equation}
    \Gamma_\mathrm{a}[\omega] = -1 + \frac{\kappa^2/2 - i\kappa\left({\omega} + \delta_\mathrm{a} \right)} {\kappa^2/4 + \delta_\mathrm{a}^2 - \lambda^2 - {\omega}^2 - i\kappa{\omega}}\;.
\label{Gamma_a}
\end{equation}
Note that this computation is carried out in the rotating frame, hence $\omega$ is the deviation from half the pump frequency. In practice we measure $\Gamma_\mathrm{a}[\omega]$ by acquiring two PNA traces (Appendinx~\ref{sec:wiring}). The first one probes the resonator under the specified pumping conditions. The second one probes the same frequency window, with the pump off and after flux tuning the resonator out of the frequency window. We divide the first trace by this second reference trace to recover $\Gamma_\mathrm{a}[\omega]$. In that respect, $|\Gamma_\mathrm{a}[\omega]|^2$ represents the frequency dependent power reflection gain of the system. Its maximum defines the gain $G \equiv \max_{\omega}|\Gamma_\mathrm{a}[\omega]|^2$.

\begin{figure}
\includegraphics[width=\linewidth,keepaspectratio]{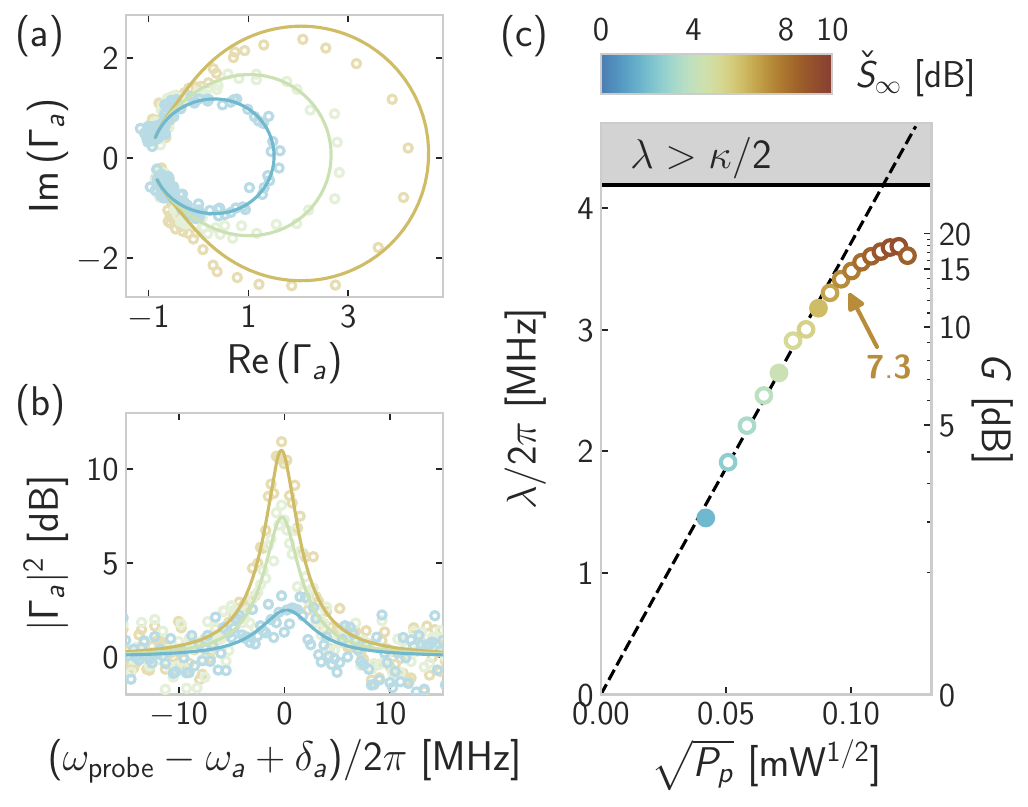}
\caption{Calibration of the two-photon pump when $\delta_\mathrm{a}=0$. (a) Complex response $\Gamma_\mathrm{a}$ of a weak reflected signal on the oscillator port for increasing steady-state squeezing (color). (b) Reflection gain magnitude in decibels (y-axis) versus detuning of the probe tone with half the pump frequency (x-axis) for increasing steady-state squeezing (color). The data in (a,b)  (open circles) are fitted to Eq.~\eqref{Gamma_a} (solid lines). (c) Two-photon pump amplitude (y-axis, left) resulting from the fits in (a,b) versus the square root of the microwave pump power at 300~K (x-axis), applied through the low-power port. The maximum power reflection gain (y-axis, right) is deduced from Eq.~\eqref{eq:gmax}, and the steady-state squeezing (colorbar, common to (a) and (b)) is deduced from Eq.~\eqref{eq:Sss}. The dashed line is a linear extrapolation of the fit before saturation. The colored arrow indicates the maximum steady-state anti-squeezing in decibels before saturation. The shaded area marks the instability region where $\lambda>\kappa/2$.}
\label{fig:cal_gain}
\end{figure}

When $|\delta_\mathrm{a}|<\kappa/2$, the reflection gain is maximized at half the pump frequency, i.e. $\omega=0$, to a value:
\begin{equation}
    G = 1 + \frac{\kappa^2 \lambda^2}{\left(\kappa^2/4+\delta_\mathrm{a}^2-\lambda^2\right)^2}\;.
    \label{eq:gmax}
\end{equation}
For a given microwave pump power, the reflection gain follows a Lorentzian lineshape around $\omega_\mathrm{p}/2$, with a full width at half maximum constrained by the gain-bandwidth product: $\Delta_{3~\text{dB}} = \kappa/\sqrt{G}$ \cite{Metelmann2022}. Knowing the characteristics of the oscillator in the absence of the pump ($\omega_\mathrm{a}, \kappa$), the two-photon pump amplitude $\lambda$ is the only free parameter when fitting the data to Eq.~\eqref{Gamma_a}. Repeating this procedure for multiple pump powers unveils the mapping between $\sqrt{P_p}$ and $\lambda$. For $\omega_\mathrm{p}=2\omega_\mathrm{a}$ this calibration is presented in Fig.~\ref{fig:cal_gain}. As the microwave pump power is increased, the fitted parameter $\lambda$ grows quadratically up to $G=14~\text{dB}$. Beyond, the fitted values keep increasing but in a slower fashion until saturation near $G=18~\text{dB}$, possibly due to a residual Kerr effect. 

\begin{figure*}
    \includegraphics[width=\linewidth,keepaspectratio]{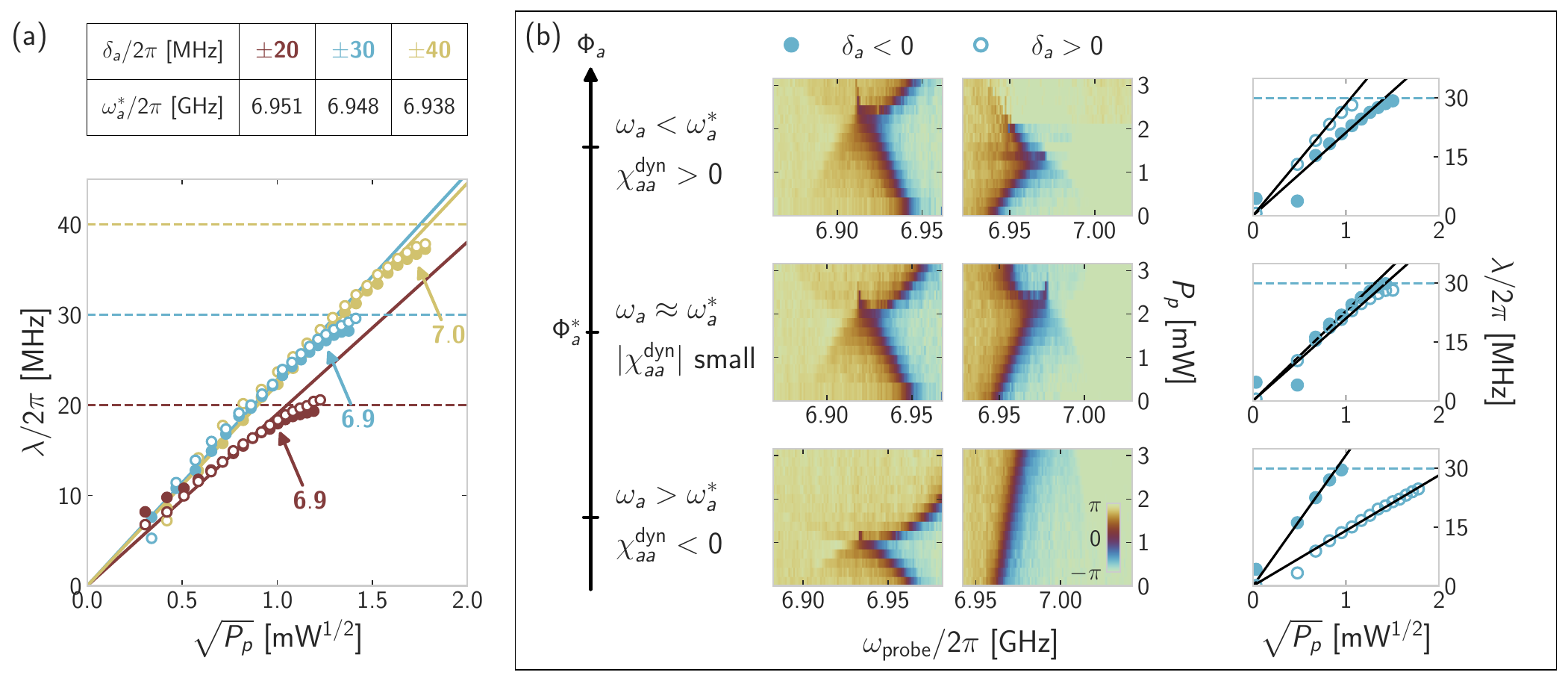}
    \caption{Calibration of the two-photon pump when $\delta_\mathrm{a}\ge\kappa/2$. (a) Two-photon pump amplitude (y-axis) as fitted from the complex response in Eq.~\eqref{Gamma_a} versus the square root of the microwave pump power at 300~K (x-axis) applied through the high-power port, for $|\delta_\mathrm{a}|/2\pi\in\{20,30,40\}$~MHz (color) and $\delta_\mathrm{a}>0$ (empty circles) or $\delta_\mathrm{a}<0$ (full circles). Colored solid lines are linear extrapolations of the fit trends before saturation. Colored arrows indicate the maximum squeezing in decibels before saturation. Colored dashed lines mark the onset of the unstable region $\lambda=|\delta_\mathrm{a}|$. The table summarizes the bare frequencies $\omega_\mathrm{a}^\ast$ of the oscillator used for the calibrations, as the ones that symmetrize the responses to a microwave pump at frequencies $\pm|\delta_\mathrm{a}|$. (b) Calibration of the dynamical Kerr-free point $\Phi_\mathrm{a}^\ast$ when $|\delta_\mathrm{a}|/2\pi=30$~MHz. From bottom to top the flux threading the SNAIL loop is increased, setting the bare oscillator frequency to 6.963~GHz (bottom), 6.949~GHz (middle) and 6.941~GHz (top). Left column: reflection phase response (color) versus input signal frequency (x-axis) and applied pump power at 300~K (y-axis) when the detuning between the bare cavity and half the pump frequency is $\delta_\mathrm{a}/2\pi=-30$~MHz. Center column: same as left for $\delta_\mathrm{a}/2\pi=+30$~MHz. The colorbar (in radians) is indicated in the bottom panel. Right column: Two-photon pump amplitude (y-axis) fitted at each pump power (x-axis) when the detuning is negative (full dots) or positive (open dots). The right and center plots of the middle panels are nearly symmetric, as demonstrated by the matching of the two types of fit on the right plot: this defines the dynamical Kerr-free point.}
    \label{fig:cal_squeeze}
\end{figure*}

When $|\delta_\mathrm{a}|\ge\kappa/2$, the reflection gain features two local maxima when $\lambda<\lambda_\mathrm{co}=\sqrt{\delta_\mathrm{a}^2-\kappa^2/4}$. These two peaks merge in the coalescent regime $\lambda \ge \lambda_\mathrm{co}$ into a single one with maximum gain given by Eq.~\eqref{eq:gmax}. Note that $\lambda_\mathrm{co}$ differs from the critical amplitude for which the gain diverges $\lambda_\mathrm{crit}=\sqrt{\delta_\mathrm{a}^2+\kappa^2/4}$. As previously, the two-photon pump amplitude $\lambda$ is the only free parameter when fitting the data to Eq.~\eqref{Gamma_a}. The calibration results are presented in Fig.~\ref{fig:cal_squeeze}(a) for $\delta_\mathrm{a}/2\pi\in\{\pm20,\pm30,\pm40\}$~MHz. We start by setting the flux at the Kerr-free point, however, when the pump is activated, the Kerr is dressed and may deviate from zero. In Fig.~\ref{fig:cal_squeeze}(b) we detail the procedure of adjusting the flux in order to reduce this dynamical Kerr effect. We display the phase response of the oscillator in the presence of an increasing pump power at $\omega_\mathrm{p}=2\omega_\mathrm{a}^0 - 2\delta_\mathrm{a}$ for $|\delta_\mathrm{a}|/2\pi=30$~MHz. Whether $\delta_\mathrm{a}$ is negative or positive, the critical value $\lambda_c$ is not reached for the same critical power $P_p^c$. Indeed, when $\delta_\mathrm{a}>0$, a positive dynamical Kerr accelerates the collapse of the oscillator signal and idler peaks. Conversely when $\delta_\mathrm{a}<0$, it slows down this process. The critical values $\lambda_c$ for each sign of the detuning only match when this spurious dynamical Kerr effect becomes negligible. When $|\delta_\mathrm{a}|/2\pi=30$~MHz and the oscillator initially sits at the Kerr-free flux point, the dynamical Kerr $\chi_\mathrm{aa}^\mathrm{dyn}$ is found to be positive (Fig.~\ref{fig:cal_squeeze}(b) top panels). Tweaking the flux bias towards higher frequencies, the two pictures can be symmetrized (middle panels), or bent in the other direction (bottom panels). All the data presented in this paper always uses the dynamical Kerr-free point associated with each value of $|\delta_\mathrm{a}|$, as recorded on the table of Fig.~\ref{fig:cal_squeeze}(a).

\subsection{Two types of squeezing}
\label{sec:ss_squeezing}

In the detuned case, the Bogoliubov transformation that diagonalizes Hamiltonian~\eqref{Harwa} defines a squeezing amplitude $S=e^{2r}$ that quantifies the anisotropy of the BO eigenstates (see Fig.~\ref{fig:fig1}). An arbitrarily large squeezing will result in anti-squeezed fluctuations in one quadrature and conversely squeezed fluctuations in the other, with no saturation. Noting that $\forall x \in (-1,1),\; \mathrm{tanh}^{-1}(x) = \frac{1}{2}\big(\mathrm{ln}(1+x) - \mathrm{ln}(1-x)\big)$, it is instructive to write the squeezing amplitude as:
\begin{equation}
    S=\sqrt{\frac{\delta_\mathrm{a}+\lambda}{\delta_\mathrm{a}-\lambda}}\;.
\end{equation}
Yet, this Bogoliubov transformation is only valid in the detuned case. In the resonant case, the two photon pump is no longer balanced by the detuning, but rather by dissipation. The oscillator reaches a steady-state, and from its quadrature statistical fluctuations we may define another squeezing parameter \cite{Gardiner2004}. Steady-state observables can be computed analytically by solving the Lindblad master equation: $\partial_t\bm{\rho}=-\frac{i}{\hbar}\left[ \bm{\mathcal{H}}_\textbf{ph}, \bm{\rho}\right] + \kappa\mathcal{D}[\ba]\bm{\rho}$. Defining the oscillator quadratures as $\bm{X_\theta}=(\ba e^{-i\theta} + \bad e^{i\theta})/2$ and $\bm{P_\theta}=(\ba e^{-i\theta} - \bad e^{i\theta})/2i$, we find for $\delta_\mathrm{a}=0$ (see Ref.~\cite{Villiers2023} for the general case):
\begin{subequations}
    \begin{align}
        \braket{\bad\ba}_\infty &= \frac{1}{2}\frac{\lambda^2}{\kappa^2/4-\lambda^2}\;, \label{eq:mean_occup} \\
        \braket{\ba^2}_\infty &= \frac{1}{2}\frac{i\lambda\kappa/2}{\kappa^2/4-\lambda^2} \;,
    \end{align}
\end{subequations}
and:
\begin{subequations}
    \begin{align}
        \braket{\bm{X_\theta}^2}_\infty &= \frac{1}{4} \frac{\kappa^2/4 + \lambda(\kappa/2)\sin 2\theta}{\kappa^2/4 - \lambda^2}\;, \\
        \braket{\bm{P_\theta}^2}_\infty &= \frac{1}{4} \frac{\kappa^2/4 - \lambda(\kappa/2)\sin 2\theta}{\kappa^2/4 - \lambda^2} \;,
    \end{align}
\end{subequations}
where $\braket{\bm{\cdot}}_\infty=\mathrm{Tr}(\bm{\cdot}\bm{\rho_\infty})$ and $\partial_t\bm{\rho_\infty}=0$. When $\lambda=0$ we recover the isotropic vacuum field fluctuation: $\forall\theta$, $\braket{\bm{X_\theta}^2}_\infty=\braket{\bm{X}^2}_\mathrm{vac}=1/4$ (same for $\bm{P_\theta}$). For $\lambda>0$, the steady-state is squeezed along $\bm{P_{\pi/4}}$ and anti-squeezed along $\bm{X_{\pi/4}}$. We define the steady-state squeezing and anti-squeezing amplitudes as:
\begin{subequations}
    \begin{align}
        S_\infty &\equiv \frac{\braket{\bm{P_{\pi/4}}^2}_\infty}{\braket{\bm{P^2}}_\mathrm{vac}} = \frac{\kappa/2}{\kappa/2+\lambda}\;, \\
        \check{S}_\infty &\equiv \frac{\braket{\bm{X_{\pi/4}}^2}_\infty}{\braket{\bm{X^2}}_\mathrm{vac}} = \frac{\kappa/2}{\kappa/2-\lambda}\;.
        \label{eq:Sss}
    \end{align}
\end{subequations}
In the large-gain limit $\lambda\rightarrow\kappa/2$, the steady-state squeezing saturates to 1/2 while the anti-squeezing grows indefinitely. This is the metric that we chose to compare the resonant case with the BO regime and its eigenstate squeezing.

Finally, in the large-gain limit one finds $\check{S}_\infty \sim \sqrt{G}$, as illustrated on Fig.~\ref{fig:cal_gain}. Combining Eqs.~\eqref{eq:gmax} (at $\delta_\mathrm{a}=0$) and \eqref{eq:mean_occup}, we find the relation $\braket{\bad\ba}_\infty=(\sqrt{G}-1)/4$. Thus, regarding Fig.~\ref{fig:TRMsqueeze}, we can estimate the mean occupancy of the resonantly squeezed oscillator at $\check{S}_\infty=8$~dB to be of approximately 1.3 photons.

\FloatBarrier
\section{Transmon characteristics and calibrations}
\label{sec:transmon}
The transmon is a superconducting qubit design featuring a Josephson element with energy $E_J$, shunted by a large capacitor with charging energy $E_C$, in the regime where $E_J\gg E_C$. It is well described by the three-level Hamiltonian:
\begin{equation}
    \bm{\mathcal{H}}_\textbf{t} = \sum_{i\in\{g,e,f\}} \hbar\omega_i \ket{i}\bra{i} \;,
\label{Hb}
\end{equation}
where $\ket{g}$, $\ket{e}$, $\ket{f}$ denote its three lowest energy states. The $\ket{g}$ and $\ket{e}$ states define the qubit states, with transition frequency $\omega_\mathrm{q}\equiv \omega_e-\omega_e = \frac{1}{\hbar}\left( \sqrt{8E_JE_C}- E_C\right)$. Single-photon excitations to the $\ket{f}$ state are detuned from the qubit transition by the anharmonicity $\chi_\mathrm{q}\approx-E_C/\hbar$ such that: $\omega_{ef}=\omega_\mathrm{q}+\chi_\mathrm{q}$.

\subsection{Single-tone spectroscopy}
\label{sec:transmon_spec}
\begin{figure}[]
\includegraphics[width=\linewidth,keepaspectratio]{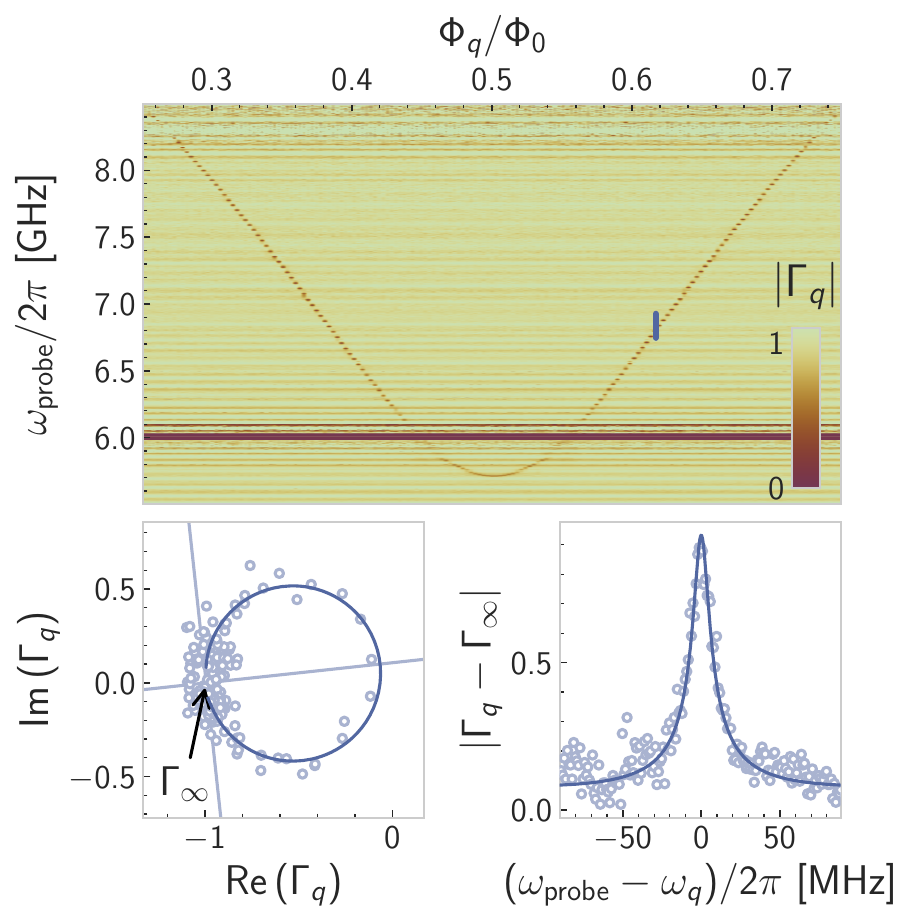}
\caption{Transmon spectroscopy versus flux. Top: Amplitude $|\Gamma_\mathrm{q}|$ of a weak reflected signal on the transmon port (color) versus the frequency of the probe tone (y-axis) and the flux threading the SQUID loop (x-axis) in unit of the flux quantum $\Phi_0=h/2e$. The detection bandwidth is interrupted by the TWPA dispersive feature around 6~GHz, and cropped by the amplifier roll-off above 8.2~GHz. Bottom: Cut of the previous map along the blue line, where $\omega_\mathrm{q}/2\pi=6.837$~GHz. Left: Complex amplitude of the reflected signal. The data is normalized to a reference background so that the accumulation point $\Gamma_\infty$ is located at coordinates (1,0). Right: Amplitude of the reflected signal referenced to the accumulation point. Fitted response (line) is overlaid to the data (circles). Light blue lines on the left plot show the symmetric axis of the circular trajectory, and its perpendicular at the accumulation point, obviously tilted in the complex frame.}
\label{fig:TRM_phi}
\end{figure}

In the present experiment, the transmon Josephson element is a SQUID. Controlling the flux $\Phi_q$ threading the SQUID loop lets us tune the transmon resonant frequency. Moreover, the transmon is strongly coupled to a microwave feedline. Photon leakage through this port dominates over every other relaxation channel. This feature was chosen to mimick the small relaxation time of typical mesoscopic qubits \cite{Cottet2017}, and also to let us record the reflection spectrum of the transmon directly, without relying on an extra readout mode (see Fig.~\ref{fig:TRM_phi}). Specifically, in the case where the transmon mode is populated with much less than 1 photon, the complex amplitude of a weak reflected signal on its input port writes:
\begin{equation}
    \Gamma_\mathrm{q}(\omega) = -1+\frac{\gamma_1}{\gamma_t/2 - i(\omega-\omega_\mathrm{q})}\;,
\label{eq:spectro_trm}
\end{equation}
where $\gamma_1$ is the qubit relaxation rate, dominated by the coupling to its feedline, and $\gamma_t=\gamma_1+2\gamma_\phi$ is the total linewidth of the transmon spectral line. Pure dephasing acts at a rate $\gamma_\phi$. The latter equation describes a circular trajectory in the complex plane, symmetric about the real axis, with an accumulation point $\Gamma_\infty=-1$. In principle, the reflection spectroscopy of such a system can distinguish the coupling rate to its feedline (here, $\gamma_1$) from the other contributions to the total linewidth (here, $2\gamma_\phi$). Using Eq.~\eqref{eq:spectro_trm} to fit the data presented in Fig.~\ref{fig:TRM_phi} (bottom panels, fit not shown) would yield $\gamma_1/2\pi=5.0$~MHz and $\gamma_\phi/2\pi=2.2$~MHz, thus placing the system near critical coupling $\gamma_1=\gamma_t/2$. However, in this very regime, fitting both rates is prone to errors due to imperfections of the experimental setup \cite{Rieger2022}. These imperfections can lead to deviations from the canonical spectroscopic response, such as tilted circles in the complex plane. It turns out that such tilts are present in the data. As a consequence, we renounce on fitting $\gamma_1$ and $\gamma_\phi$ separately. Rather, we employ a fit function representing circles with any orientation in the complex plane, thus sensitive to $\gamma_t$ only (see Fig.~\ref{fig:TRM_phi} bottom panels, blue lines). This procedure lets us fit the total linewidth of the transmon line reliably and accurately.

\subsection{Two-tone spectroscopy}
\label{sec:transmon_twotone}

\begin{figure}[]
    \includegraphics[width=\linewidth,keepaspectratio]{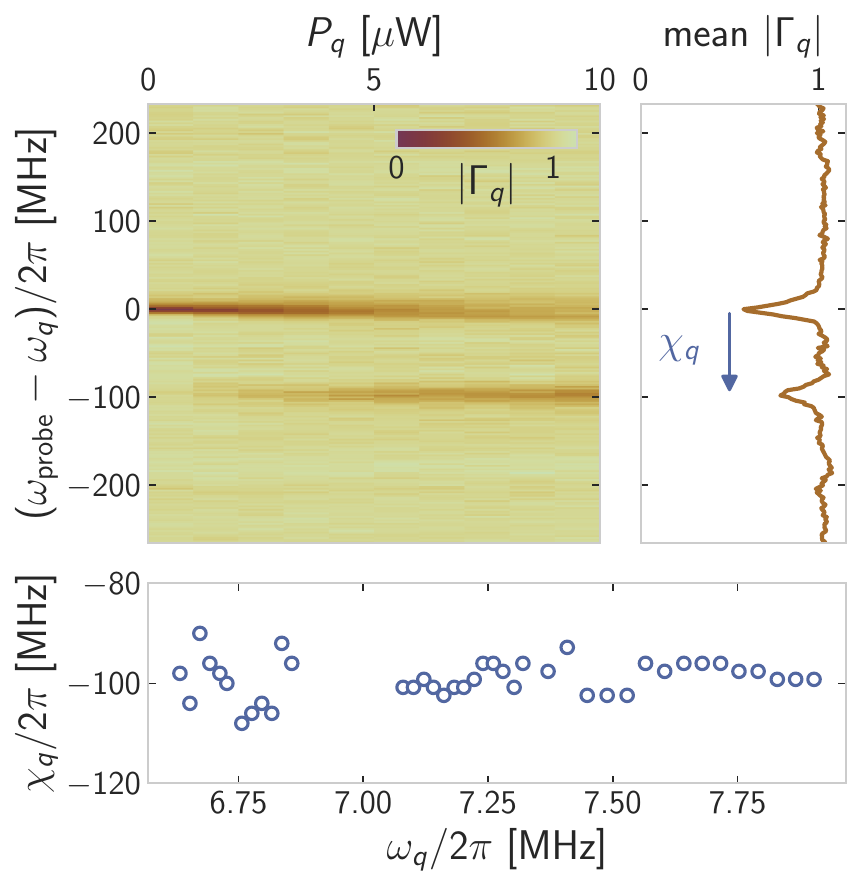}
    \caption{Two-tone spectroscopy of the transmon. Top: (Left) Amplitude $\Gamma_\mathrm{q}$ (color) of a weak reflected signal with frequency $\omega_\text{probe}$ on the transmon port , while the transmon is driven on resonance at $\omega_\mathrm{q}$ with a microwave tone of power $P_q$ at 300~K (x-axis). (Right) Average signal over all powers $P_q$. The distance between the g-e transition peak at resonance and the emerging e-f transition peak yields the anharmonicity $\chi_{q}$. Bottom: Anharmonicity versus resonant frequency of the transmon, as extracted by repeating the experiment shown in the top panels at mutliple flux points $\Phi_q$.}
    \label{fig:two-tone}
    \end{figure}

So far, the anharmonicity of the transmon has been disregarded. Unlike the previous discussion, driving the transmon with higher powers unravels its multi-level structure. We reveal transmon states beyond the qubit manifold by performing a two-tone spectroscopy, saturating the g-e transition with a resonant microwave drive, and then probing the transmon with a weak tone (see Fig. \ref{fig:two-tone}). Due to the finite occupation of the $\ket{e}$ state provided by the saturation drive, the e-f transition can be revealed by the weak tone. Note that the spectroscopic tone is about 5000 times less powerful than the saturation one. We repeat the experiment at multiple flux points, thus varying the qubit frequency. The fitted anharmonicty fluctuates around $-100$~MHz, the value predicted by electromagnetic simulations of the transmon design.

\subsection{Transmon-oscillator resonant coupling}
\label{sec:transmon_resonant}
\begin{figure}[]
\includegraphics[width=\linewidth,keepaspectratio]{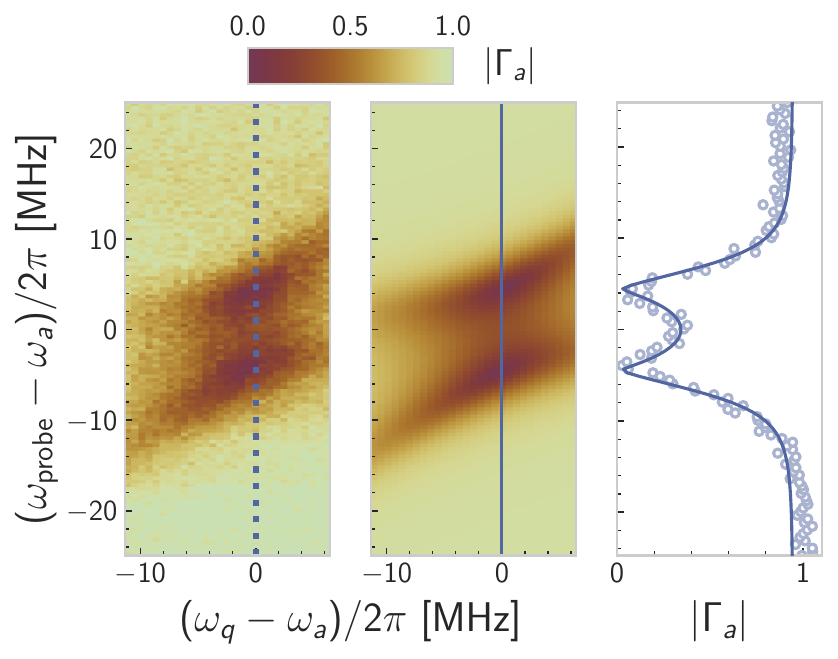}
\caption{Anti-crossing of the transmon and resonator spectral lines. Left: Amplitude $\Gamma_\mathrm{a}$ (color) of a weak reflected signal with frequency $\omega_\text{probe}$ on the oscillator port (y-axis), versus the qubit frequency (x-axis), referenced to the oscillator frequency at the Kerr-free point. Center: amplitude of the fitted response with the coupling rate $g$ as the only fitting parameter. Right: line-cut of the measured spectrum (open circles) and the fitted response (solid line) at resonance as indicated by the dotted (full) line in the left (middle) plot.}
\label{fig:anticrossing}
\end{figure}

Making the most of the wide tunability range of both the transmon and oscillator frequencies, we can study their interaction in different detuning regimes. We begin with the resonant case (see Fig.~\ref{fig:anticrossing}). Setting the oscillator to its Kerr-free flux point, we record its reflection spectrum as the transmon frequency is swept accross. From input-output theory we expect the following response:
\begin{equation}
    \Gamma_\mathrm{a}(\omega) = -1 + \dfrac{\kappa}{\dfrac{\kappa}{2} - i(\omega-\omega_\mathrm{a}) + \dfrac{g^2}{\frac{\gamma_t}{2} - i(\omega-\omega_\mathrm{q})}}\;,
\end{equation}
where $g$ is the resonant coupling amplitude. Having previously calibrated the decay rates of the oscillator ($\kappa/2\pi=8.7$~MHz) and the transmon ($\gamma_t/2\pi=8.0$~MHz at the oscillator Kerr-free point), the recorded map can be fitted using $g$ as the only fitting parameter. When the transmon and the oscillator are on resonance, the oscillator spectrum displays a partially resolved splitting. Indeed, the coupling amplitude $g/2\pi=6.1$~MHz is smaller than the decay rates of both modes, thus placing the system just below the strong resonant coupling regime. 

\subsection{Transmon-oscillator dispersive coupling and photon number calibration}
\label{sec:disp_cal}
Next we turn to the characterization of the coupling in the dispersive limit: $|\omega_\mathrm{q}-\omega_\mathrm{a}|\gg g$. Following Ref.~\cite{Schuster2005, Gambetta2006}, the dispersive interaction can be revealed through the measurement of the AC-Stark shift $\Delta\omega_\mathrm{q}$ and induced dephasing $\Delta\gamma_\phi$ of the qubit upon coherent driving of the oscillator. Specifically, a coherent drive at frequency $\omega_d$ and power $P_\mathrm{drive}$ stabilizes a coherent field $\alpha_{g,e}$ in the cavity, whether the qubit is in $\ket{g}$ or $\ket{e}$, such that:
\begin{equation}
    \alpha_{g,e}=\frac{-i\kappa/2}{\kappa/2 + i(\omega_\mathrm{a}-\omega_d\mp\chi/2)}\sqrt{\frac{P_\mathrm{drive}}{P_0}}\;,
\end{equation}
where $P_0$ is the drive power maintaining one photon in the oscillator (regardless of the qubit state since $|\chi|\ll\kappa$). 
%The qubit spectral properties are subsequently modified through $\Delta\omega_\mathrm{q} = \Re(\chi\alpha_g^\ast\alpha_e)$ and $\Delta\gamma_\phi = -\Im(\chi\alpha_g^\ast\alpha_e)$, where $\chi$ is the dispersive interaction amplitude. 
Subsequently, the finite occupation of the oscillator shifts the qubit frequency by $\Delta\omega_\mathrm{q} = \mathrm{Re}(\chi\alpha_g^\ast\alpha_e)$, where $\chi$ is the dispersive interaction amplitude. Moreover, the occupation number of the coherent field follows Poisson statistics,  leading to an induced dephasing of the qubit: $\Delta\gamma_\phi = -\mathrm{Im}(\chi\alpha_g^\ast\alpha_e)$.
In the weak-dispersive limit $|\chi|\ll\kappa$ and resonant driving, these formula simplify to: $\Delta\omega_\mathrm{q}=\chi \bar{n}_\mathrm{d}$ and $\Delta\gamma_\phi=2\chi^2 \bar{n}_\mathrm{d}/\kappa$, where $\bar{n}_\mathrm{d}=P_\mathrm{drive}/P_0$ is the mean photon number injected by the coherent drive in the oscillator. Both the dispersive coupling $\chi$ and the photon-number calibration $P_0$ can be extracted from the joint fitting of the AC-Stark shift and induced dephasing with the applied drive power.

\begin{figure}[]
\includegraphics[width=\linewidth,keepaspectratio]{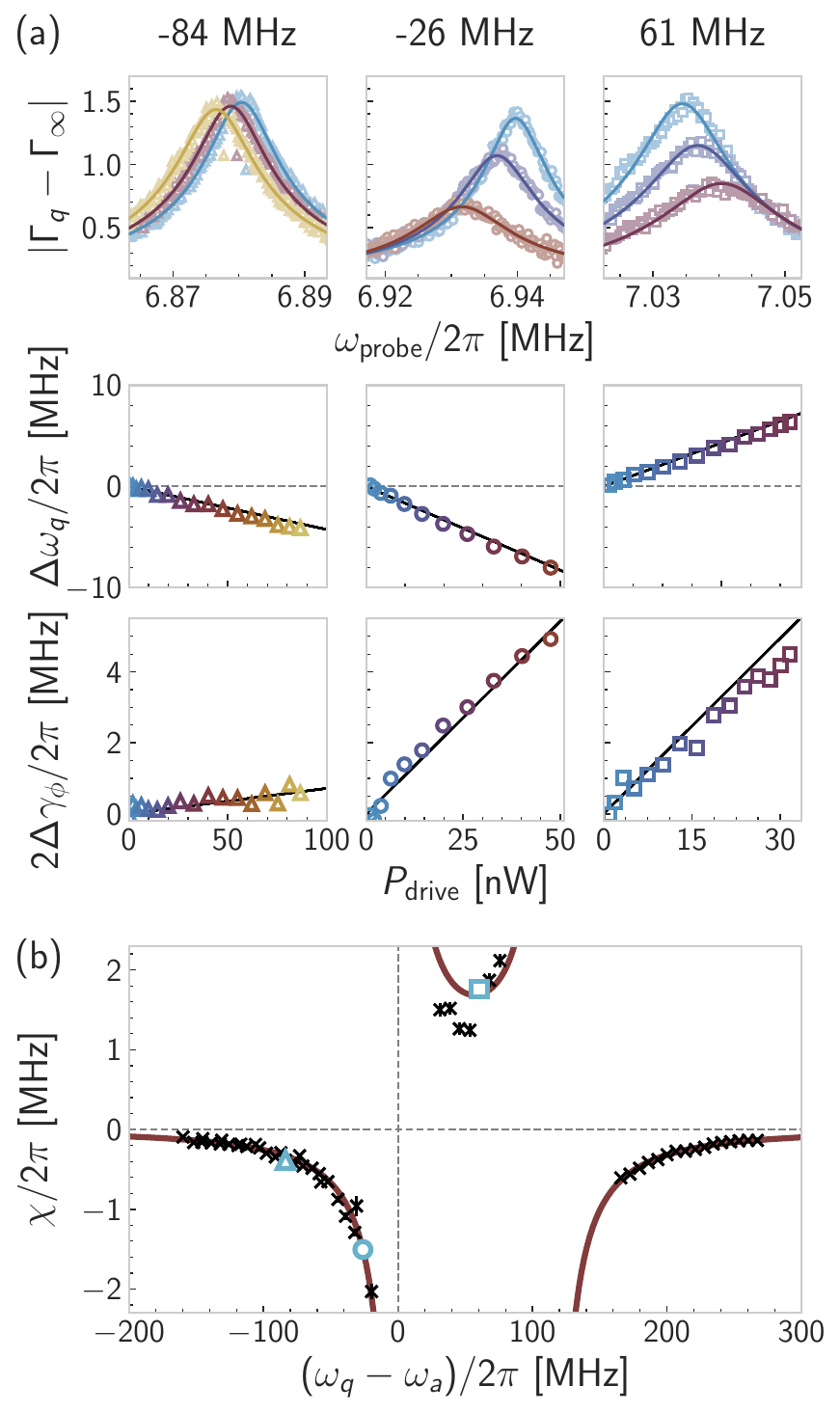}
\caption{Straddling regime and oscillator photon number calibration. (a) Top: Amplitude (y-axis) of a weak reflected signal with frequency $\omega_\mathrm{probe}$ (x-axis) on the transmon port in the presence of a resonant microwave drive on the oscillator with increasing power (color) and transmon-oscillator detuning of $-84$~MHz (left, triangles), $-26$~MHz (center, circles) and $61$~MHz (right, squares). The induced dephasing and AC-Stark shift on the transmon are extracted by fitting the complex data (open symbols) to circles in the complex plane (solid lines). Middle: AC-Stark shift of the transmon (y-axis) versus applied power on the oscillator at 300~K (x-axis, color). Bottom: Linewidth broadening of the transmon spectral line (y-axis) versus applied power on the oscillator at 300~K (x-axis, color). Linear fits (black lines) of the AC-Stark shift and the induced dephasing versus drive power yield the dispersive interaction amplitude $\chi$ for each transmon-oscillator detuning, and the photon number calibration $P_0$  (see text). (b) Dispersive coupling (y-axis) versus transmon-oscillator detuning (x-axis). The data (black crosses) are fitted with the analytical result (bordeaux solid line) accounting for the transmon $f$ state (see Eq.~\eqref{eq:straddle}). Blue open symbols mark the detunings used in (a).}
\label{fig:straddle}
\end{figure}

This procedure is presented in Fig.~\ref{fig:straddle} for multiple qubit frequencies $\omega_\mathrm{q}$, while the oscillator sits at its Kerr-free point and is driven at resonance $\omega_d=\omega_\mathrm{a}$. For each value of $\omega_\mathrm{q}$, we measure the qubit AC-Stark shift $\Delta\omega_\mathrm{q}$ and measurement induced dephasing $\Delta\gamma_\phi$ as a function of the drive power $P_\text{drive}$ on the oscillator. We fit this entire dataset to the above formula keeping as free parameters: $\chi$ at every qubit frequency, and a single power calibration $P_0$. We find $P_0=8.1\pm0.3$~nW. Also, the evolution of $\chi$ with qubit-oscillator detuning clearly displays the straddling regime: when the oscillator frequency lies between the $g\text{-}e$ and $e\text{-}g$ transitions, virtual transitions to the $\ket{f}$ state strongly affects the dispersive interaction strength \cite{Koch2007}. We fit the extracted $\chi$ versus $\omega_\mathrm{q}$ to the analytical result accounting for the transmon $\ket{f}$ state:
\begin{equation}
    \chi = \frac{2g^2}{\omega_\mathrm{q}-\omega_\mathrm{a}}\frac{\chi_\mathrm{q}}{\omega_\mathrm{q}-\omega_\mathrm{a}+\chi_\mathrm{q}}\;.
\label{eq:straddle}
\end{equation}
Keeping as free parameters $g$ and $\chi_\mathrm{q}$, we find $g/2\pi=4.9$~MHz and $E_C/h=-\chi_\mathrm{q}/2\pi= 114$~MHz, which are close to the values extracted from the anti-crossing and two-tone spectroscopy described in previous sections.

Finally, we perform another photon-number calibration, this time with an oscillator drive $300$~MHz above resonance. We set the qubit-oscillator detuning to $-100$~MHz, for which the dispersive interaction strength is $\chi/2\pi=-250$~kHz. Then, the photon number calibration is extracted from a fit of the qubit AC-Stark shift versus the applied detuned microwave power on the oscillator. We find $P_0^{+300}=230$~$\mu$W, which we use to estimate the oscillator occupancy in the Kerr spectroscopy experiment (Fig.\ref{fig:SPA_phi}).

% \FloatBarrier
\section{Dispersive interaction of a qubit and a resonantly squeezed oscillator}
\label{sec:disp_amp}
In this appendix, we review the modification of the qubit spectral properties when the SNAIL-resonator is pumped at the degenerate parametric resonance $\omega_\mathrm{p}=2\omega_\mathrm{a}$. While the evolution of the measurement induced dephasing as a function of the cavity gain was covered extensively in Refs.~\cite{Eddins2019, Levitan2015}, analysis of the concurrent frequency shift and thus the dispersive interaction strenght was not addressed.

Starting from the system Hamiltonian~\eqref{eq:H_JC}, we can write a SW transformation that leaves invariant the bare oscillator part, including the two-photon pump. Its generator reads:
\begin{equation}
\begin{aligned}
    \bm{\mathcal{S}} =&\; \frac{g}{\delta_\mathrm{q}-\delta_\mathrm{a}}\dfrac{1}{1-\dfrac{\lambda^2}{\delta_\mathrm{q}^2-\delta_\mathrm{a}^2}} \ba\sgp - \mathrm{h.c.} \\
    &+ \frac{\lambda g}{\delta_\mathrm{q}^2-\delta_\mathrm{a}^2}\dfrac{1}{1-\dfrac{\lambda^2}{\delta_\mathrm{q}^2-\delta_\mathrm{a}^2}} \ba\sgm  - \mathrm{h.c.}\;. \label{eq:ac_gain}
\end{aligned}
\end{equation}
While this change of frame was used in the BO regime $\kappa/2\ll\lambda<|\delta_\mathrm{a}|$ in Ref.~\cite{Shani2022}, here we focus on the usual amplifier regime $|\delta_\mathrm{a}|<\kappa/2$. Note that when $\delta_\mathrm{a}=0$, the qubit frequency in the rotating frame corresponds to the qubit-oscillator detuning. This will be the regime of interest for the remainder of this appendix. Introducing the dimensionless parameter $\eta$ such that: $g, \: \kappa < \eta \times |\delta_\mathrm{q}|$, and under the assumption that $\eta \ll 1$, Hamiltonian~\eqref{eq:H_JC} reads in the transformed basis $\ba \rightarrow e^{-\bm{\mathcal{S}}} \ba e^{\bm{\mathcal{S}}}$, $\sgz \rightarrow e^{-\bm{\mathcal{S}}} \sgz e^{\bm{\mathcal{S}}}$ up to second order in $\eta$:
\begin{align}
    \bm{\mathcal{H}}_\textbf{q-ph}/\hbar =\:& - \frac{\lambda}{2}\left( \ba^2 + \badd \right)  \\
    &+ \left[\delta_\mathrm{q} + \chi\left(\bad\ba + \frac{1}{2}\right) \right] \frac{\sgz}{2} \;, \nonumber
\end{align}
where $\chi = 2g^2/\delta_\mathrm{q}$ is the bare dispersive interaction parameter. Corrections to $\chi$ occur at order four in $\eta$. 

\begin{figure}[]
    \includegraphics[width=\linewidth,keepaspectratio]{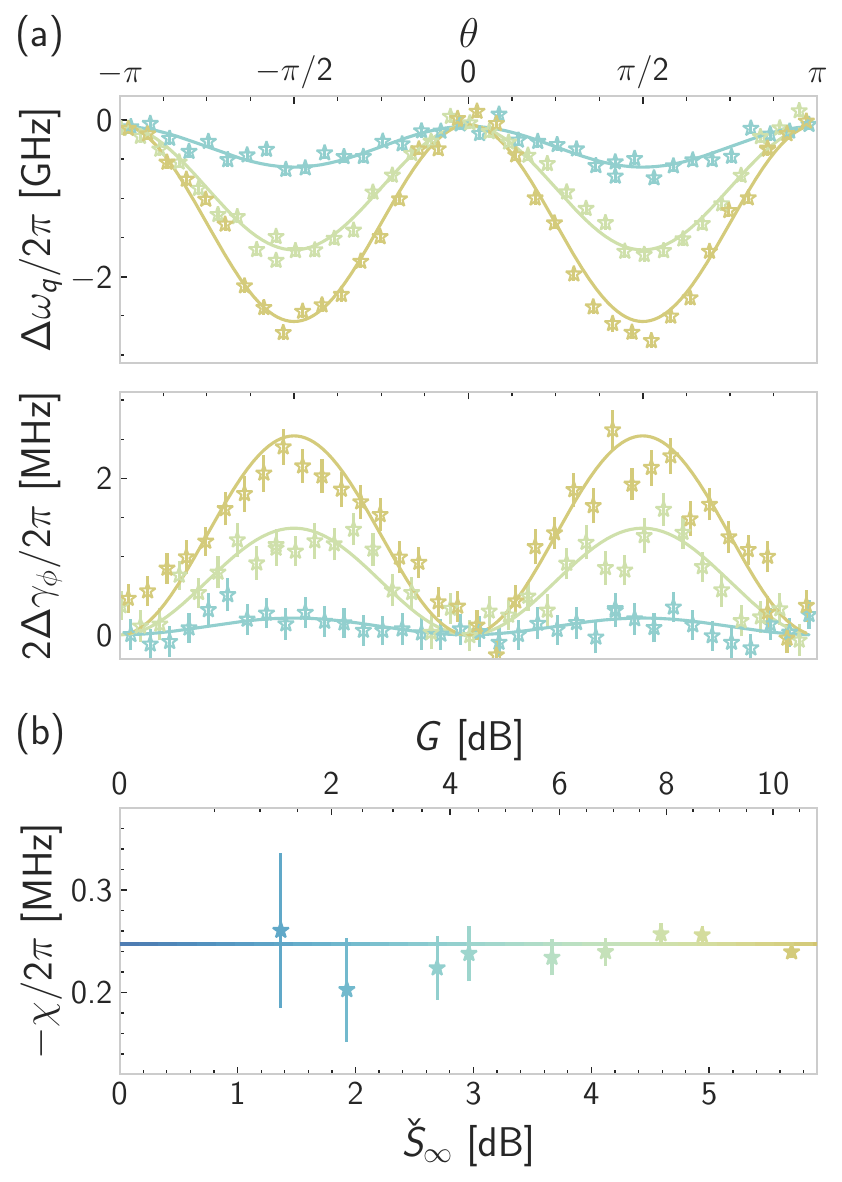}% Here is how to import EPS art
    \caption{Dispersive interaction of the qubit with a resonantly squeezed oscillator $\delta_\mathrm{a}=0$. (a) Top: AC-Stark shift $\Delta\omega_\mathrm{q}[\lambda, \bar{n}_\mathrm{d}]$ (y-axis) versus drive phase (x-axis) for various oscillator steady-state anti-squeezing (color) and a fixed drive amplitude. Bottom: same for linewidth broadening $2\Delta\gamma_\phi[\lambda, \bar{n}_\mathrm{d}]$ (y-axis). Data points (stars) are extracted in a procedure akin to Fig.~\ref{fig:straddle}. Solid lines are fits to Eq.~\eqref{eq:ac_gain} and Eq.~(4) of Ref.~\cite{Eddins2019}, yielding $\chi$ for each steady-state anti-squeezing value. (b) Dispersive interaction strength (y-axis) versus steady-state anti-squeezing (x-axis bottom and color) and amplifier gain (x-axis top). The solid line marks the theoretical value that is expected to be independent of anti-squeezing up to second order in $\kappa/\delta_\mathrm{q}$.}
    \label{fig:chi_G}
\end{figure}

Like in the previous part, $\chi$ can be inferred from the joint measurement of the AC-Stark shift and linewidth broadening of the qubit in the presence of a microwave drive, resonant with the oscillator. Following the derivation of Appendix~\ref{sec:theory_mid}, the dressing of the qubit spectral features are deduced from the steady-state properties of the oscillator, to lowest order in $\chi/\kappa$. The oscillator dynamics is governed by the Lindblad master equation $\partial_t\bm{\rho}=-\frac{i}{\hbar}\left[ \bm{\mathcal{H}}_\textbf{ph}+\bm{\mathcal{H}}_\textbf{drive}, \bm{\rho}\right] + \kappa\mathcal{D}[\ba]\bm{\rho}$, where in the rotating frame $\bm{\mathcal{H}}_\textbf{drive}/\hbar=(\varepsilon_d/2)\ba + (\varepsilon^\ast_d/2)\bad$. Since $\delta_\mathrm{a}=0$, the drive frequency is commensurate with the pump frequency, and their relative phase is expected to modify the system response \cite{Eddins2019}. The drive complex amplitude is defined as $\varepsilon_d=|\varepsilon_d|e^{i(\theta-\pi/4)}$, so that when $\theta=0$ the in-phase component of the drive lies along the squeezed quadrature of the oscillator (Appendix~\ref{sec:ss_squeezing}). Defining the mean occupation of the oscillator in the steady-state $\bar{n}_\mathrm{a}=\mathrm{Tr}(\bad\ba\bm{\rho_\infty})$ with $\partial_t\bm{\rho_\infty}=0$, the qubit frequency shift reads $\Delta\omega_\mathrm{q} \equiv \chi\bar{n}_\mathrm{a}=\Delta\omega_\mathrm{q}[\lambda] +\Delta\omega_\mathrm{q}[\lambda, \bar{n}_\mathrm{d}]$ where:
\begin{subequations}
\begin{align}
    \Delta\omega_\mathrm{q}[\lambda] &= \frac{1}{2}\frac{\lambda^2}{\kappa^2/4-\lambda^2}\chi  \;,\\
    \Delta\omega_\mathrm{q}[\lambda, \bar{n}_\mathrm{d}] &= \frac{\kappa^2}{4}\frac{\kappa^2/4 + \lambda^2 - \lambda\kappa\cos 2\theta }{(\kappa^2/4 - \lambda^2)^2 }\bar{n}_\mathrm{d} \chi \;,
\end{align}
\end{subequations}
and $\bar{n}_\mathrm{d}=|\varepsilon_d|^2/\kappa^2$ is the mean photon number injected by the coherent drive with the pump off. The first term is the modified Lamb shift measured in Fig.~\ref{fig:TRMsqueeze}. The second one is the modified AC-Stark shift. Together with the phase dependent induced dephasing (see Eq. (4) of Ref.~\cite{Eddins2019}), we can infer the dispersive interaction strength from a joint fitting of the qubit spectral features versus the drive phase, at a fixed drive amplitude. This procedure is presented in Fig.~\ref{fig:chi_G} for various oscillator steady-state anti-squeezing, and a calibrated injected photon number $\bar{n}_\mathrm{d}=0.8$ (Appendix~\ref{sec:disp_cal}). We find a dispersive interaction strength independent of the steady-state anti-squeezing over the whole measurement range, with an amplitude matching the transmon version of $\chi$ (see Eq.~\eqref{eq:straddle}).

\section{Dispersive interaction of a transmon and squeezed photons}
\label{sec:theory_trm}
\begin{figure}[]
    \includegraphics[width=\linewidth,keepaspectratio]{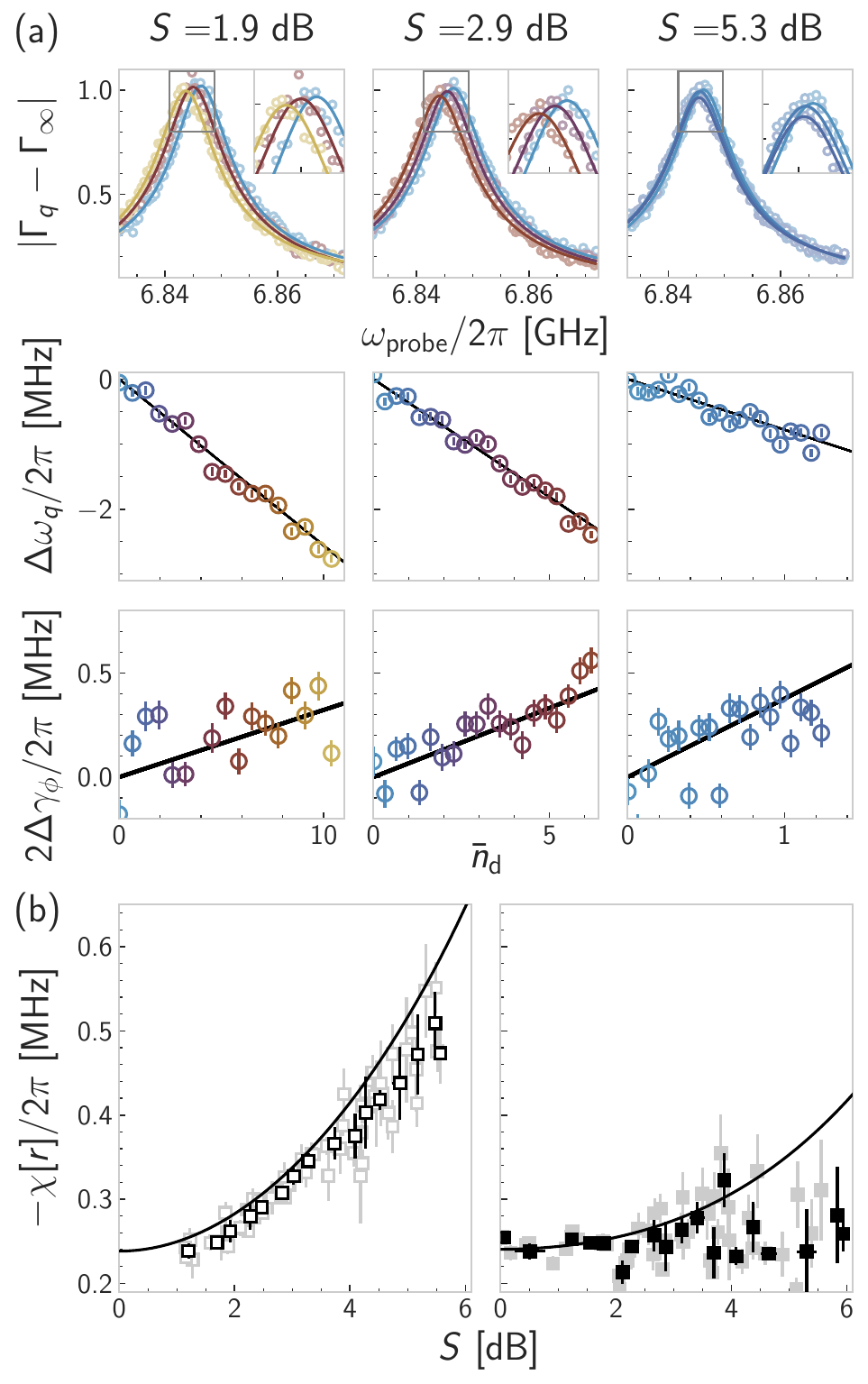}
    \caption{Dispersive interaction of the transmon with the BO when $|\delta_\mathrm{a}|/2\pi=20$~MHz. (a) Top: Amplitude (y-axis) of a weak reflected signal with frequency $\omega_\mathrm{probe}$ (x-axis) on the transmon port in the presence of a resonant microwave drive on the BO injecting an increasing number of photons (color), and BO squeezing of $1.9$~dB (left), $3.6$~dB (center) and $5.3$~dB (right) with $\delta_\mathrm{a}>0$. The induced dephasing and AC-Stark shift are extracted by fitting the complex data (open symbols) to circles in the complex plane (solid lines). Insets: zoom on the resonance as indicated by the grey box. Middle: AC-Stark shift of the transmon $\Delta\omega_\mathrm{q}[r, \bar{n}_\mathrm{d}]$ (y-axis) versus number of injected photons (x-axis, color). Bottom: same for linewidth broadening $2\Delta\gamma_\phi[r, \bar{n}_\mathrm{d}]$. Linear fits (black lines) of the AC-Stark shift (Eq.~\eqref{eq:ac_squeeze_trm}) and the induced dephasing (Eq.~\eqref{eq:mid_squeeze_trm}) versus calibrated injected photon number yield the dispersive interaction amplitude $\chi[r]$ for each BO squeezing. (b) Left: dispersive interaction strength (y-axis) versus BO squeezing (x-axis) for $\delta_\mathrm{a}>0$. The full data record (lightgrey squares, 72 points) is coarse-grained (black squares), and compared with the analytical result with no fit parameters (black line, Eq.~\eqref{eq:chitrm}). Right: same for $\delta_\mathrm{a}<0$.}
    \label{fig:chi_vs_S_detail}
\end{figure}
%\FloatBarrier

In this appendix we extend the results of Appendix~\ref{sec:theory} obtained for a two-level system to the higher energy levels of the transmon, beyond the qubit manifold. 
%Even though we solely considered its three lowest energy levels in Appendix~\ref{sec:transmon}, in this part we do not specify any cutoff for the transmon ladder. 
Following Ref.~\cite{Koch2007}, the transmon-BO Hamiltonian reads in the Bogoliubov basis and under the RWA:
\begin{equation}
\begin{aligned}
    \bm{\mathcal{H}}_\textbf{t-ph}/\hbar &= \Omega_\mathrm{a}[r] \bald\bal + \sum_k \delta_k \ket{k}\bra{k} \\
    &+ \sum_k g_{k,k+1}\cosh r \left(\bal \ket{k+1}\bra{k} + \mathrm{h.c.} \right) \\
    &+ \sum_k g_{k,k+1}\sinh r \left(\bal \ket{k}\bra{k+1} + \mathrm{h.c.} \right)
\end{aligned}
\label{eq:Htph_rwa}
\end{equation}
where $\delta_k = \omega_k - k\times\frac{\omega_\mathrm{p}}{2}$ and $g_{k,k+1} \approx g\sqrt{k+1}$. We neglect multi-photon transitions in the transmon spectrum. Indeed, despite the multi-level structure of the transmon, in the limit $E_J\gg E_C$ selection rules forbid photo-assisted transitions between non-neighboring energy levels. Moving on with the dispersive transformation, the generator of the SW unitary reads:
\begin{equation}
\begin{aligned}
    \bm{\mathcal{S}} = &\sum_k \frac{g_{k,k+1}\cosh r}{\delta_{k,k+1}-\Omega_\mathrm{a}[r]} \left(\bal \ket{k+1}\bra{k} - \mathrm{h.c.} \right) \\
    &-\sum_k \frac{g_{k,k+1}\sinh r}{\delta_{k,k+1}+\Omega_\mathrm{a}[r]} \left(\bal \ket{k}\bra{k+1} - \mathrm{h.c.} \right)
\end{aligned}
\end{equation}
where $\delta_{k,k+1}=\delta_{k+1}-\delta_k$. We introduce a dimensionless parameter $\eta$ such that: $\forall k, \; g_{k,k+1}e^r < \eta \times \mathrm{min}(|\delta_{k,k+1}\pm\Omega_\mathrm{a}[r]|)$. This dispersive transformation requires that all the allowed transmon transitions are detuned from the BO signal and idler frequencies. This regime is safely maintained in our experiment. In the transformed basis $\bal\rightarrow e^{-\bm{\mathcal{S}}}\bal e^{\bm{\mathcal{S}}}$, and $\forall k,\:\ket{k} \rightarrow e^{-\bm{\mathcal{S}}} \ket{k}$, the restriction of Hamiltonian~\eqref{eq:Htph_rwa} to the two lowest energy transmon levels reads at second-order in $\eta$:
\begin{align}
    \bm{\mathcal{H}}_\textbf{t-ph}/\hbar =& \left( \Omega_\mathrm{a}[r] + \Omega_\mathrm{a}^{(2)}[r] \right)\bald \bald + \left( \delta_\mathrm{q} + \delta_\mathrm{q}^{(2)}[r] \right)\frac{\sgz}{2} \nonumber \\
     &+ \chi[r] \bald\bal \frac{\sgz}{2}\;,
\end{align}
where we introduced the qubit-manifold spin operator $\sgz = \ket{e}\bra{e} - \ket{g}\bra{g}$, and:
\begin{subequations}
\begin{align}
    \Omega_\mathrm{a}^{(2)}[r] =\;& -\frac{g^2\cosh^2r}{\Delta[r]} -\frac{g^2\sinh^2r}{\Sigma[r]}\;, \\
    \delta_\mathrm{q}^{(2)}[r] =\;& \frac{g^2\cosh^2r}{\Delta[r]} + \frac{g^2\sinh^2r}{\Sigma[r]}\frac{\chi_\mathrm{q}-\Sigma[r]}{\chi_\mathrm{q}+\Sigma[r]} \;, \\
    \chi[r] =\;& \frac{2g^2}{\Delta[r]} \frac{\chi_\mathrm{q}}{\chi_\mathrm{q}+\Delta[r]} \cosh^2 r  \label{eq:chitrm} \\
    &+ \frac{2g^2}{\Sigma[r]} \frac{\chi_\mathrm{q}}{\chi_\mathrm{q}+\Sigma[r]} \sinh^2 r\;. \nonumber
\end{align}
\end{subequations}
One can check that in the limit $|\chi_\mathrm{q}|\gg|\Delta[r]|,\:|\Sigma[r]|$ we recover $\delta_\mathrm{q}^{(2)}[r] \approx \chi[r]/2$, as customary for a dispersively coupled qubit.

Finally, the frequency shift of a transmon dispersively coupled to a driven BO reads $\Delta\omega_\mathrm{q}=\Delta\omega_\mathrm{q}[r] + \Delta\omega_\mathrm{q}[r,\bar{n}_\mathrm{d}]$ where: 
\begin{subequations}
\begin{align}
    \Delta\omega_\mathrm{q}[r] &= \delta_\mathrm{q}^{(2)}[r] + \chi[r]\sinh^2r\;, \\
    \Delta\omega_\mathrm{q}[r,\bar{n}_\mathrm{d}] &= \chi[r] \bar{n}_\mathrm{d}\cosh^2 r \;. \label{eq:ac_squeeze_trm}
\end{align}
\end{subequations}
Similarly the linewidth broadening of the transmon qubit transition reads $\Delta\gamma_\phi = \Delta\gamma_\phi[r] + \Delta\gamma_\phi[r,\bar{n}_\mathrm{d}]$ where:
\begin{subequations}
\begin{align}
    \Delta\gamma_\phi[r] &= \frac{\chi^2[r]}{\kappa} \sinh^2 r \left(1+\sinh^2r\right)\;, \\ 
    \Delta\gamma_\phi[r,\bar{n}_\mathrm{d}] &= \frac{2\chi^2[r]}{\kappa} \left(1 + 2\sinh^2r\right)\bar{n}_\mathrm{d} \cosh^2r\;. \label{eq:mid_squeeze_trm}
\end{align}
\end{subequations}
These are the transmon version of Eqs.~\eqref{eq:shift_qubit} and \eqref{eq:broad_qubit}, used to fit the data in Figs.~\ref{fig:TRMsqueeze} and \ref{fig:chi_vs_S}. The experimental procedure used to measure the dispersive interaction strength when $|\delta_\mathrm{a}|/2\pi=20$~MHz is detailed in Fig.~\ref{fig:chi_vs_S_detail}. We repeat it for $|\delta_\mathrm{a}|/2\pi\in\{30,40\}$~MHz, which yields the data of Fig.~\ref{fig:chi_vs_S}.

\FloatBarrier
\bibliography{villiers2024}% Produces the bibliography via BibTeX.

\end{document}